\documentstyle[seceq,supplement,wrapft, epsf]{ptptex}
   \notypesetlogo 
   \pubinfo{No. 133, pp. 53-84, 1999}

   \def\lesssim{
   \mathrel{\hbox{\rlap{\hbox{\lower4pt\hbox{$\sim$}}}\hbox{$<$}}}}
   \def\gtrsim{
   \mathrel{\hbox{\rlap{\hbox{\lower4pt\hbox{$\sim$}}}\hbox{$>$}}}}

   \newcommand{\der}{\partial}
   \newcommand{\abs}[1]{\left|{#1}\right|}

   \newcommand{\bk}[1]{\langle{#1}\rangle}
   \newcommand{\am}{\!'}
   \newcommand{\as}{\!''}
   
\begin{document}
\markboth{ K.~Umetsu, M.~Tada and T.~Futamase}
{Cluster Mass Reconstruction by Weak Shear Field}

\title{\bf Cluster Mass Reconstruction by a Weak Shear Field}

\author{Keiichi {\sc Umetsu}
\footnote{E-mail address: keiichi@astr.tohoku.ac.jp},\\
  Masashi {\sc Tada}
\footnote{E-mail address: tada@astr.tohoku.ac.jp}  
and   Toshifumi {\sc Futamase}
\footnote{E-mail address: tof@astr.tohoku.ac.jp}
}
\inst{Department of Astronomy, 
Graduate School of Science,
Tohoku University, \\
Sendai  
980-8578, Japan}
\abst{
The tidal gravitational field of galaxy clusters  
causes a coherent distortion of the images
 of background sources. 
 Since the distribution of image distortions, namely the shear field,
 traces the local gravitational potential of a deflector, 
 it can be used to reconstruct
 the two-dimensional mass distribution of clusters of galaxies.
 Moreover, the shear field can 
 provide unique information on the redshift distribution of
 high-redshift galaxies.
 In this review we summarize recently-developed parameter-free 
 methods of cluster-mass
 reconstruction based on the shear field, and we apply a mass-reconstruction
 method to the cluster Abell 370 at redshift 0.375. 
}

\maketitle

\section{Introduction}
                                   
Clusters of galaxies are the most massive gravitational bound systems in the
universe and therefore contain crucial information on the origin of 
structure formation and on cosmology.\cite{HA91,Rich92,BLD95} \
For example, statistical studies of
the cluster mass distribution and its evolution give us 
information  
on the distribution of dark matter and 
on the cosmological density parameter $\Omega_0$.

To this time,
various
methods have been used to determine the mass
distributions in clusters of galaxies, such as:
a dynamical method in which 
the velocity dispersion 
of member galaxies obtained from optical observations
is used on the basis of the virial theorem,\cite{BT87} \   
and a method based on 
diffuse X-ray emission from the hot ($\sim$ several keV) intra-cluster medium
(ICM) that
traces the gravitational potential of the cluster.\cite{Sarazin86}  \
These two methods, however, are based on
strong assumptions which are not fully justified. 
In applying the dynamical method, we must require that clusters 
are in virial equilibrium. However, this requirement 
is not always satisfied, because
the typical
dynamical time scale of a galaxy cluster is not 
much shorter than the Hubble
time $H_0^{-1}$, at least for a high-$\Omega_0$ universe, and  
the existence of substructures observed in some clusters
indicate that they are not relaxed systems. 
Moreover, projection
effects of substructures  and the anisotropy of galaxy orbits in clusters 
may 
cause substantial errors in
mass determinations by this method.
On the other hand, X-ray analysis
requires the assumption that the ICM is in hydrostatic equilibrium with
the gravitational potential and also requires spherical symmetry for the
mass profile of the cluster.   
These two methods are thus very sensitive to the physical and the dynamical
state.

Another approach to determine the mass distribution of clusters is to
make use of 
gravitational lensing, which can in principle be used to determine
 the projected mass
distribution of the lens object directly,
independent of the physical state and
the nature of the matter content.
The light rays from background sources are
deflected by clusters of galaxies, 
and the strength of the deflection depends on
the total mass and the mass distribution of the deflector.

In the case of
strong lensing, which occurs near the cluster core,
the light-ray bending leads to
the formation of {\it arcs} in cluster fields. 
If we assume spherical symmetry for the
cluster mass distribution, the distance of an arc to the cluster center
yields an estimate for the mass enclosed by a circle traced by the arc.
An arc therefore 
places strong constraints on the cluster-mass distribution,
especially in the central region of the cluster.

On the other hand, 
weakly-distorted images of 
background galaxies---so called {\it arclets}---can be observed to
much larger angular separations from the cluster center, 
and thus they
provides valuable information concerning 
cluster-mass distributions out to large radii.\cite{Web85} \
Tyson, Valdes and Wenk\cite{TVW90} were the first to detect the
weakly-distorted
coherent images of galaxies behind two rich clusters.
Kochanek\cite{Kocha90} and
Miralda-Escud$\acute{e}$\cite{Mir91}
attempted to fit parameterized mass
profiles of the clusters to the observed distortion fields.
Kneib et al.\cite{Kneib93} placed strong constraints on 
the bimodal mass profile for
the cluster Abell 370 using one giant luminous arc 
(hereafter referred to as GLA) and
multiple images. Moreover, it has been shown that arc statistics are very 
sensitive to cosmological parameters, such as the cosmological
constant $\Lambda$.\cite{WuM96,FHH98} \
Bonnet, Mellier and Fort\cite{BMF94} measured the coherent gravitational
shear induced by the cluster Cl 0024+16 
out to $3\;h_{50}^{-1}$ Mpc, where $h_{50}$ is the Hubble parameter in
units of 50 km s$^{-1}$ Mpc$^{-1}$.
It was first found by Kaiser and Squires\cite{KS93} that 
the distortion field can be used for a parameter-free reconstruction of the
two-dimensional mass density of a cluster in the weak-lensing regime.
That is, the two-dimensional  mass density of a deflector can be expressed
as a convolution integral of the gravitational shear---
which can be obtained from the image distortions
of background galaxies in the weak-lensing limit---with a known kernel. 
The reconstruction formula derived by Kaiser and Squires was 
then generalized to include the strong-lensing regime\cite{K95,ScSe95,SeSc95} 
and modified to remove the artificial boundary
effects,\cite{K95,Bart95,Sc95,SeSc96,SqK96} \
which
arise in the resulting mass map obtained with the original Kaiser and Squires 
mass-reconstruction technique.
These mass-reconstruction methods have been used to obtain the mass
distributions of clusters in recent years (e.g.,
Refs.~\citen{F94}$\sim$~\citen{Se96}. \ 
%

In the present paper, we summarize 
the recent progress on the theoretical and the observational front
in this field. There have been several reviews of this
subject.\cite{FM94,Mell96,SchBart97} \ 
We shall try to be {\it self-contained} and {\it pedagogical} 
so that the reader
 can perform the cluster-mass reconstruction by reading this paper.  
The remainder of the paper is organized as follows: In {\S} 2 we
present basic equations and concepts of gravitational lensing.  
In {\S} 3 we introduce the definition of the image shape, 
and describe how the local observables are related to the local
properties of the lensing clusters. 
In Section {\S} we review some mass-reconstruction schemes
that are based solely on image distortions. 
As mentioned earlier, applications of the
reconstruction formula derived by Kaiser and Squires to real data
involve several
difficulties. Here we describe the difficulties 
which one encounters in
applying 
the original reconstruction method to real clusters 
and how such difficulties can be overcome. 
Section 5 summarizes 
the observational studies of galaxy clusters based on the shear analysis. 
In {\S} 6
we apply a mass-reconstruction method to the cluster Abell 370 to demonstrate
the feasibility of the shear-based analysis. 
There we employ a new 
method to break the mass degeneracy associated with the mass
reconstructions based solely on image shapes.
Finally, we conclude with a summary 
in {\S} 7.


\section{Gravitational Lensing}
In this section we summarize the basic equations and concepts involved
in 
gravitational lensing which we shall need later.
For general treatments, see 
Refs.~\citen{SEF92,BN92,NBart96}.

\subsection{Basic relations}
The gravitational field of a deflector changes the 
source position
$\vec{\beta}$ to the apparent position $\vec {\theta}$ according to the
lens equation. 
Let $\Sigma(\vec{\theta})$ be the surface mass density of a deflector.
The lens equation is written in terms of the
two-dimensional effective lensing potential $\psi(\vec{\theta})$ in the form
\begin{equation}
\vec{\beta}=\vec{\theta}-\vec{\nabla}_{\!\theta}\psi(\vec\theta). 
\label{lenseq}
\end{equation}
Here $\psi(\vec{\theta})$ satisfies 
the two-dimensional Poison
equation,
\begin{equation}
\triangle_{\theta} \psi(\vec{\theta})=2\kappa(\vec{\theta}), \label{Poison}
\end{equation}
where $\kappa$ is the convergence  
defined by
\begin{equation}
 \kappa(\vec{\theta}) := \frac{\Sigma(\vec{\theta})}{\Sigma_{\rm cr}},
\end{equation}
with the critical surface-mass density 
\begin{equation}
\Sigma_{\rm cr}:=\frac{c^2 D_{\rm s}}{4\pi G D_{\rm d} D_{\rm ds}}.
\end{equation}
Here $D_{\rm d}$, $D_{\rm s}$ and $D_{\rm ds}$ 
are the angular-diameter distances from the observer to the deflector, 
from the observer to the source and from the deflector to the source,
respectively. 
For a fixed lens redshift, we see that $\psi(\vec{\theta})$ depends on 
the source redshift 
through the distance ratio $D_{\rm ds}/D_{\rm s}$ in $\Sigma_{\rm cr}$.

Since $\psi(\vec {\theta})$ satisfies the two-dimensional Poison
equation  (\ref{Poison}), it can be written
in terms of $\kappa$ as
\begin{equation}
 \psi(\vec {\theta})=\frac{1}{\pi}\int d^2 \theta'\,
   \ln|\vec {\theta}-\vec {\theta}'|\,\kappa(\vec{\theta}').
\end{equation}
We then introduce the components of the shear,
\begin{equation}
\gamma_1:=\frac{1}{2}(\psi,_{11}-\psi,_{22})\,\,\,;\,\,\,\gamma_2:=\psi,_{12},
\label{gg1}
\end{equation}
where an index $i$ $(i=1,2)$ preceded by a comma denotes partial 
derivatives with respect to $\theta^i$.
It is useful to combine the shear components to define the complex shear,
\begin{equation}
 \gamma:=\gamma_1+i\gamma_2\equiv |\gamma|\exp(2i\phi).
\label{gg2}
\end{equation}
We further define the {\it tangential shear} $\gamma_{\rm
t}(\vec{\theta},\vec{\theta}')$ relative to the point $\vec{\theta}'$ by
\begin{eqnarray}
\gamma_{\rm t}(\vec{\theta};\vec{\theta}')&:=&
-\left(
\gamma_1(\vec{\theta})
\cos\left[2\varphi(\vec{\theta},\vec{\theta}')\right]
+\gamma_2(\vec{\theta})
\sin\left[2\varphi(\vec{\theta},\vec{\theta}')\right] 
\right)\nonumber\\
&=&
\Re\left[
\gamma\,{\rm e}^{-2i(\varphi+\frac{\pi}{2})}
\right]
=|\gamma|\cos\left[2(\phi-\varphi-\frac{\pi}{2})\right],\label{tshear}
\end{eqnarray}
where  $\Re(z)$ denotes the real part of a complex number $z$,
and $\varphi(\vec{\theta},\vec{\theta}')$ is defined by
\begin{equation}
\varphi(\vec{\theta},\vec{\theta}'):=
\arctan\left(\frac{\theta_2-\theta_2'}{\theta_1-\theta_1'}\right).
\end{equation}
From the definition, 
$\gamma(\vec{\theta})$ is a linear combination
of the second derivatives of $\psi(\vec{\theta})$, 
so that the complex shear can be expressed by the convolutional
integral of $\kappa$ with a known kernel:
\begin{eqnarray}
\gamma(\vec{\theta})&=&
\left(\frac{\der_1^2-\der_2^2}{2}+i\der_1\der_2\right)\,\psi(\vec{\theta})\\
&=&\frac{1}{\pi}\int d^2 \theta'\,
   \mbox{$\cal D$}(\vec{\theta}-\vec{\theta}')\,\kappa(\vec{\theta}').
 \label{gamma}
\end{eqnarray} 
Here $\der_i:=\der/\der\theta^i\, (i=1,2)$ and 
$\mbox{$\cal D$}(\vec{\theta})$ is the complex kernel,
\begin{eqnarray}
\mbox{$\cal D$}(\vec{\theta})&:=&
\left(\frac{\der_1^2-\der_2^2}{2}+i\der_1\der_2\right)\,\ln|\vec{\theta}|
\nonumber\\
&=&
\frac{\theta_2^2-\theta_1^2-2i\theta_1\theta_2}{|\vec{\theta}|^4}
=\frac{-1}{(\theta_1-i\theta_2)^2}. \label{kernelD}
\end{eqnarray}
We note that there exists a {\it global} transformation that leaves 
$\gamma(\vec{\theta})$ invariant such that
\begin{equation}
\kappa(\vec{\theta}) \to \kappa(\vec{\theta})+\kappa_0,
\end{equation}
where $\kappa_0$ is an arbitrary constant. 
We see in {\S} 4 
that this degeneracy
leads to an ambiguity in the  mass distribution predicted by the 
reconstruction technique based solely on 
the shear field.
\subsection{Magnification and image distortion}

The local properties of the lens equation are described by its Jacobian
matrix:
\begin{equation}
 {\cal A}(\vec{\theta}):=\frac{\der \vec{\beta}}{\der \vec{\theta}}
=\left(
  \begin{array}{@{\,}cc@{\,}}
 1-\psi,_{11} & -\psi,_{12} \\
 -\psi,_{12}         & 1-\psi,_{22}
  \end{array}
\right). \label{Jacob}
\end{equation}
%
From Eqs. (\ref{Poison}), (\ref{gg1}) and (\ref{gg2}), the 
Jacobian matrix ${\cal A}$ can be rewritten 
as 
\begin{eqnarray}
 {\cal A}(\vec{\theta})
&=&
\left(
  \begin{array}{@{\,}cc@{\,}}
 1-\kappa-\gamma_1 & -\gamma_2 \\
 -\gamma_2         & 1-\kappa+\gamma_1
  \end{array}
\right) \nonumber \\
&=& (1-\kappa)
\left(
  \begin{array}{@{\,}cc@{\,}}
 1 & 0 \\
 0 & 1
  \end{array}
\right)
-\abs{\gamma}
\left(
  \begin{array}{@{\,}cc@{\,}}
 \cos(2\phi) &  \sin(2\phi) \\
 \sin(2\phi) & -\cos(2\phi)
  \end{array}
\right),\label{matrixA2}
\end{eqnarray}
and it has  two eigenvalues 
\begin{equation} 
\Lambda_{\pm}:=1-\kappa \pm |\gamma|.
\end{equation}
We see from Eq. (\ref{matrixA2}) that the convergence $\kappa$ is 
the trace part, and
the complex shear $\gamma$ is the trace-free part of the Jacobian matrix
${\cal A}(\vec \theta)$. 
The convergence term alone in Eq. (\ref{matrixA2}) causes an 
isotropic focusing of the light ray, while the shear term induces
 anisotropy in the lens mapping.
The quantity $|\gamma|=(\gamma_1^2+\gamma_2^2)^{1/2}$ is the amplitude
of the shear, and the phase $\phi$ is the position angle of the
eigenvector that belongs to 
the eigenvalue $\Lambda_{-}=1-\kappa-|\gamma|$.

The magnification of an image is given by the convergence and the shear 
as follows:
\begin{equation}
\mu(\vec{\theta}):=
\left|
\mbox{det}\left(
\frac{\der \vec{\theta}}{\der \vec{\beta}}
\right)
\right|
=\frac{1}{
|\mbox{det}{\cal A}|}
=\frac{1}{|(1-\kappa)^2-|\gamma|^2|}. \label{mag}
\end{equation}

\begin{wrapfigure}{r}{6.6cm}
\epsfxsize=7cm
\centerline{\epsfbox{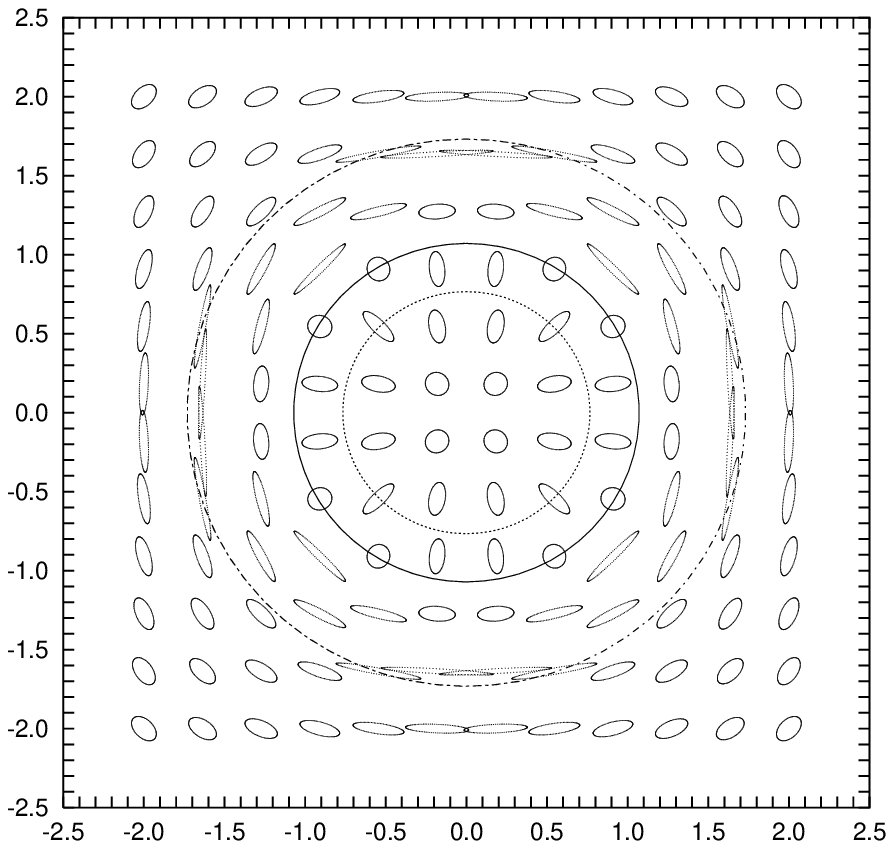}}
\caption{
Image-distortion field of intrinsically-circular sources 
for a non-singular circularly-symmetric lens.
The dotted line represents the inner critical curve,
where images are radially elongated.
The dashed line represents the outer critical curve,
 where images are tangentially elongated.  
Image distortions vanish along the curve 
$\kappa=1$ indicated by the solid line. 
The curve $\kappa=1$ lies in the odd-parity region.
}
\label{image}
\end{wrapfigure}
\noindent
The closed curves defined by
\begin{equation}
0=\mbox{det}{\cal A}(\vec{\theta})
\end{equation} 
are called
`critical curves', on which the magnification diverges.
The image plane is separated by critical curves;
the regions where $\mbox{sign}(\mbox{det}{\cal A})=+1$ and $-1$
are called {\it even}- and {\it odd}-parity regions, respectively.
%
%
An intrinsically-circular source is transformed to an
ellipse with axis ratio ($\leq 1$) of $|\Lambda_-/\Lambda_+|$ 
for $\kappa<1$ and 
$|\Lambda_+/\Lambda_-|$ for $\kappa\geq 1$ and magnified by the factor
$\mu=1/|\Lambda_-\Lambda_+|$.
The gravitational distortion vanishes along the curve defined by 
$\kappa(\vec{\theta})=1$, which lies in the odd-parity region.
%

In particular, for a non-singular  circularly-symmetric lens, 
there are two {\it circular}
critical curves: the inner critical curve defined by $0=\Lambda_+
=1-\kappa+|\gamma|$, and the outer critical curve defined by
$0=\Lambda_- = 1-\kappa-|\gamma|$.
Images close to the outer critical curve are elongated in the tangential 
direction, while images close to the inner critical curve are radially
elongated (see Fig. \ref{image}). 
From these image properties, the outer and inner
critical curves are called {\it tangential} and {\it radial},
respectively. 
%
%

\section{ Observables vs lensing properties}

As stated in {\S} 2.2, the tidal component of the gravitational field 
causes a coherent distortion of the images of background sources.
Hence, the galaxy images observed through a cluster can be used to infer 
the lensing properties of the cluster.
For this purpose, we must first quantify the shape 
of a galaxy image in terms of observable quantities.
For an elliptically-shaped source, its shape and orientation can be defined in
terms of the axis ratio and the position angle of the major axis,
respectively. In this case, an appropriate image ellipticity
can also  be defined by them.
However, observations of galaxies reveal 
quite irregular shapes which cannot be accurately
approximated by a simple ellipse, and this irregularity leads to  
serious noise in measuring the shear field. 
We must, therefore, take account of the
irregularity in quantifying a galaxy shape.

First, we define the center of a galaxy image by
\begin{equation}
\vec{\theta}_{\rm c}
:=\frac{\int d^2 \theta \,q[I(\vec{\theta})]\,\vec{\theta}}
{\int d^2 \theta \,q[I(\vec{\theta})]}, \label{galposi}
\end{equation}
where $I(\vec{\theta})$ is the surface-brightness distribution 
of the source galaxy, 
and  $q(I)$ is a weight function, which we choose appropriately.
For example, a possible choice for $q(I)$ may be the Heaviside step
function $q(I)=H(I-I_{\rm th})$, 
where $I_{\rm th}$ is a limiting surface brightness, 
and $H(x)$ is defined by 
\begin{equation}
H(x):=
\left\{
\begin{array}{@{\,}llll}
 1      &\mbox{for} & x \geq 0  \\
 0      &\mbox{for} & x < 0
\end{array}
\right. .
\end{equation}
In this case $\vec{\theta}_{\rm c}$ is the center of the area 
enclosed by a
threshold isophote $I(\vec{\theta})=I_{\rm th}$. For $q(I)=I$, 
$\vec{\theta}_{\rm c}$ is the center of light.
We then introduce the 
tensor of second brightness moments for each image:
\begin{equation}
 Q_{ij}:=\frac{\int d^2 \theta \,q[I(\vec{\theta})]\,
(\theta-\theta_{\rm c})_i(\theta-\theta_{\rm c})_j}
{\int d^2 \theta \,q[I(\vec{\theta})]} \ , \ i,j\in\{1,2\}. \label{Qij}
\end{equation}
With this definition, we can quantify the shape of a galaxy image 
using the complex
ellipticity $\epsilon=\epsilon_1+i\epsilon_2$:\cite{Sc95,SeSc97}
\begin{equation}
 \epsilon := \frac{Q_{11}-Q_{22}+2iQ_{12}}
{Q_{11}+Q_{22}+2(Q_{11}Q_{22}-Q_{12}^2)^{1/2}}
\equiv|\epsilon|\,{\rm e}^{2i\phi}
\label{eps}.
\end{equation}
If galaxy images have elliptical isophotes 
with axis ratio $r$ ($\leq 1$), then the complex ellipticity is
$\epsilon=(1-r)/(1+r)\exp (2i\phi)$,  
where the phase $\phi\in[0,\pi)$ is the
position angle of the major axis. 

Next, we consider the intrinsic (unlensed) surface-brightness
distribution of the galaxy image $I_{\rm s}(\vec{\beta})$, where
the source position $\vec{\beta}$ is related to the image position
$\vec{\theta}$ through the lens equation (\ref{lenseq}). 
Liouville's theorem tells us that 
gravitational lensing conserves surface brightness, and therefore
we can set
\begin{equation}
I_{\rm s}(\vec{\beta})=I(\vec{\theta}).
\end{equation}
We then define the tensor of second brightness moments ${Q_{\rm s}}_{ij}$ 
of an unlensed source as
\begin{equation}
 {Q_{\rm s}}_{ij}:=\frac{\int d^2 \beta \,q[I_{\rm s}(\vec{\beta})]\,
(\beta-\beta_{\rm c})_i(\beta-\beta_{\rm c})_j}
{\int d^2 \beta \,q[I_{\rm s}(\vec{\beta})]}
\,\,\,, \,\,\,i,j\in \{1,2\}, \label{Qsij}
\end{equation}
where $\vec{\beta}_{\rm c}$ is the angular position of the 
center of the unlensed source.
If we assume that source images are smaller than the angular scale 
where 
the lensing properties change, then we can locally linearize the lens 
equation to obtain
\begin{equation}
(\beta-\beta_{\rm c})_i={\cal A}_{ij}(\vec{\theta}_{\rm c})
\,(\theta-\theta_{\rm c})^j.
\label{linear}
\end{equation}
From Eqs. (\ref{Qij}), (\ref{Qsij}) and (\ref{linear}), we find that the
tensors of second brightness moments of the source and image are
related according to
\begin{equation}
 Q_{\rm s}={\cal A}Q{\cal A}^{T}={\cal A}Q{\cal A},
\end{equation}
where ${\cal A}\equiv{\cal A}(\vec{\theta}_{\rm c})$ is the Jacobian matrix 
of the lens
equation at angular position $\vec{\theta}_{\rm c}$.
If we further define
the complex ellipticities $\epsilon_{\rm s}$ of the unlensed source 
in the same way as in Eq. (\ref{eps}), 
the transformation between source and image
ellipticity is given by
\begin{equation}
\epsilon(\epsilon_{\rm s},g)=
\left\{
\begin{array}{@{\,}lll}
 (\epsilon_s+g)/(1+g^{*}\epsilon_{\rm s}) 
&\mbox{for} & |g| \leq 1  ,\\
 (1+g \epsilon_{\rm s}^*)/(\epsilon_{\rm s}^*+g^{*})  
&\mbox{for}& |g| > 1 ,
\end{array}
\right. \label{eg1}
\end{equation}
where $g(\vec\theta)$ is the {\it reduced shear} defined by
\begin{equation}
g:=\gamma / (1-\kappa),
\end{equation}
and the asterisk denotes complex conjugation.
Since $\mbox{det}{\cal A}=(1-\kappa)^2\,(1-|g|^2)$, the condition $|g|<
1$ ($|g|\geq 1$) is equivalent to the condition $\mbox{det}{\cal A}>0$
($\mbox{det}{\cal A}\leq 0$).

Hence, the lensing properties are related to the observable
quantity $\epsilon$ only through the reduced shear $g$. 
In particular,  
the intrinsic ellipticity $\epsilon_{\rm s}$
vanishes for circular sources, and thus we have
\begin{equation}
 \epsilon=
\left\{
\begin{array}{@{\,}lll}
 g      &\mbox{for} & |g| \leq 1 , \\
 1/g^*  &\mbox{for} & |g| > 1 .
\end{array}
\right.  \label{eg2}
\end{equation}
In general, on the other hand,  
source galaxies have intrinsic ellipticities, 
so that only one image gives us no
information on the local gravitational field.
We can, however, extract 
the local lensing properties
from the observed distortion field
by assuming that source galaxies are randomly oriented. If this
assumption holds, that is, if 
$\epsilon_{\rm s}=|\epsilon_{\rm s}|\exp(2i\phi)$ 
has a random phase $\phi$, then the probability distribution function 
$p_{\epsilon_{\rm s}}(\epsilon_{\rm s})$ of 
the source ellipticity $\epsilon_{\rm s}$
in general takes the form
\begin{equation} 
p_{\epsilon_{\rm s}}(\epsilon_{\rm s})\,d^2\epsilon_{\rm s}
=p(|\epsilon_{\rm s}|)\,
|\epsilon_{\rm s}|
d|\epsilon_{\rm s}|\,
\frac{d\phi}{2\pi},
\end{equation}
where the function $p(|\epsilon_{\rm s}|)$ is normalized so that
$
1=\int_0^1 d|\epsilon_{\rm s}| \,|\epsilon_{\rm s}|\,p(|\epsilon_{\rm s}|)
$.
It has been  shown by Seitz and Schneider\cite{SeSc97} 
that the expectation value of
the $n$-th moment $\epsilon^n$ is related to the reduced shear $g$ through
\begin{eqnarray}
\bk{\epsilon^n}_{\epsilon_{\rm s }}
&:=&
\int d^2\epsilon_{\rm s}\,p_{\epsilon_{\rm s}}\,\epsilon^n(\epsilon_{\rm s};g)
\nonumber\\
&=&
\int_0^1\!d|\epsilon_{\rm s}|\,|\epsilon_{\rm s}|\,
p(|\epsilon_{\rm s}|)
\oint\!\frac{d\phi}{2\pi}\,\epsilon^n(\epsilon_{\rm s};g)\nonumber\\
&=&
\left\{
\begin{array}{@{\,}llll}
 g^n      &\mbox{for} & |g| \leq 1,  \\
 1/{g^*}^n  &\mbox{for} & |g| > 1 .
\end{array}
\right. \label{eg3}
\end{eqnarray}
Note that the expectation value $\bk{\epsilon^n}_{\epsilon_{\rm s}}$ 
does not depend on the 
source-ellipticity distribution  $p(|\epsilon_{\rm s}|)$; 
if we adopt a different definition of the image ellipticity 
(e.g., Ref.~\citen{ScSe95})
its expectation value depends on its distribution. 

In practice, however, we must replace the expectation value 
$\bk{\epsilon}_{\epsilon_{\rm s}}$ with 
the average over a local ensemble of image ellipticities, $\bar{\epsilon}$:
\begin{equation}
  \bk{\epsilon}_{\epsilon_{\rm s}}
(\vec{\theta})\approx\bar{\epsilon}(\vec\theta),
\end{equation}
where
\begin{equation}
\overline{\epsilon^n}(\vec{\theta}):=
\frac{\sum_{i=1}^{N_{\rm gal}} W(\vec{\theta}-\vec{\theta}_i)\,
        \epsilon^n(\vec{\theta}_i)}
     {\sum_{i=1}^{N_{\rm gal}} W(\vec{\theta}-\vec{\theta}_i)}. \label{epsbar}
\end{equation}
Here $N_{\rm gal}$ is the number of galaxy images, and $\vec{\theta}_i$ is the
angular position of the $i$-th galaxy ($i=1,2,\cdots,N$) defined by
Eq. (\ref{galposi}),
and $W(\vec{\theta})$ is the weight function
\begin{equation}
W(\vec{\theta})=\exp\left(-\frac{|\vec{\theta}|^2}{\Delta\theta^2}\right)
\label{weightfunc}
\end{equation}
with smoothing scale $\Delta\theta$.
This scale must be small enough so that the
lensing properties can be assumed constant over the effective 
smoothing disk 
of area $\sim \pi\Delta\theta^2$ but large enough so that the
smoothing disk contains a sufficient number of galaxies to suppress the
random noise.\cite{SeSc95,SeSc96} 

We can thus make use of the smoothed ellipticity field 
$\bar{\epsilon}(\vec{\theta})$ as a direct estimator 
for $g(\vec{\theta})$ or $1/g^*(\vec{\theta})$.
In the case of weak lensing ($\kappa\ll 1$ and $|\gamma|\ll 1$, i.e., 
$|g|\ll 1$), 
we have
\begin{equation}
\gamma\approx g\approx \bar{\epsilon}.
\end{equation}
In this limit, the shear $\gamma$ is a direct observable.

\section{Cluster mass reconstruction by weak shear field}

In this section we review several cluster-inversion methods 
based on the shear analysis.
The point of this study is that the two-dimensional mass distribution of 
a cluster can be directly 
obtained only from the observed shear field, and that we
need not  assume {\it a priori} a mass profile of a cluster.
In {\S} 4.1 we review the original mass-reconstruction method
developed by Kaiser and Squires\cite{KS93} and 
summarize several difficulties
we encounter in a practical application of this method to real data.
In {\S} 4.2-4 we describe how these difficulties can be overcome.
In {\S} 4.5 a method to infer the total mass within a circular
aperture is described. 

\subsection{Kaiser and Squires algorithm---linear inversion
Formula---}

Since both $\kappa$ and $\gamma$ are linear combinations of the
second derivatives of $\psi$, using the relation in Fourier space,
one can obtain an expression for $\kappa$ in terms of the complex shear
 $\gamma$; that is, the relation (\ref{gamma}) can be inverted.\cite{KS93} 

To see this, we express the convergence $\kappa(\vec{\theta})$ 
by its Fourier transform $\hat{\kappa}(\vec{k})$ as
\begin{equation}
\kappa(\vec{\theta})=\frac{1}{(2\pi)^2}\int d^2 k\,\hat{\kappa}(\vec{k})\,
\exp(i\vec{k}\cdot\vec{\theta}) \label{Fourier}.
\end{equation}
In Fourier space, Eq. (\ref{gamma}) reads
\begin{equation}
\hat{\gamma}(\vec{k})=\frac{1}{\pi}\,\hat{\kappa}(\vec{k})\,
\hat{\mbox{$\cal D$}}(\vec{k}), \label{kapgamma}
\end{equation}
where 
$\hat{\kappa}(\vec{k})$ and $\hat{\mbox{$\cal D$}}(\vec{k})$ 
are the Fourier transforms of 
$\kappa(\vec{\theta})$ and $\mbox{$\cal D$}(\vec{\theta})$, respectively,
and they
are defined in the same way as
in Eq. (\ref{Fourier}). 
The Fourier transform of the complex kernel 
$\mbox{$\cal D$}(\vec{\theta})$ defined by (\ref{kernelD}) is 
\begin{equation}
\hat{\mbox{$\cal D$}}(\vec{k})=\pi\,\frac{k_1^2-k_2^2+2ik_1k_2}{|\vec{k}|^2}.
\label{kernelDc}
\end{equation}
From Eq. (\ref{kernelDc}), we find that 
$\hat{\mbox{$\cal D$}} \hat{\mbox{$\cal D$}}^*=\pi^2$,
implying 
\begin{equation}
\hat{\mbox{$\cal D$}}^{-1}=\pi^{-2}\,\hat{\mbox{$\cal
D$}}^*. \label{inverse}
\end{equation}
Inserting Eq. (\ref{inverse}) into Eq. (\ref{kapgamma}) yields
\begin{equation}
\hat{\kappa}(\vec{k})=\frac{1}{\pi}\,\hat{\gamma}(\vec{k})\hat{\mbox{$\cal D$}}^*(\vec{k}).
\end{equation}
This relation can be converted into the relation in the  real
$\vec{\theta}$-space as follows:
\begin{eqnarray}
\kappa(\vec\theta)-\kappa_0
 &=&\frac{1}{\pi}\int d^2 \theta'\,
   \mbox{$\cal D$}^{*}(\vec\theta-\vec\theta')\,\gamma(\vec\theta') \nonumber\\
  &=&\frac{1}{\pi}\int d^2 \theta'\,
   \Re[\mbox{$\cal D$}^{*}(\vec\theta-\vec\theta')\,\gamma(\vec\theta')]
\label{KS}.
\end{eqnarray}
Here the constant $\kappa_0$ is inserted into the right-hand side of 
Eq. (\ref{KS}) because
adding a uniform mass sheet does not change the shear 
(see Eq. (\ref{gamma})). In terms of the tangential shear 
$\gamma_{\rm t}$ defined by Eq. (\ref{tshear}), 
the inversion equation takes the following form: 
\begin{equation}
\kappa(\vec\theta)-\kappa_0=
\frac{1}{\pi}\int d^2 \theta'\,
\frac{\gamma_{\rm t}(\vec{\theta'};\vec{\theta})}
{|\vec{\theta}-\vec{\theta}'|^2}.
\end{equation}  
Thus we obtain an expression for the convergence $\kappa(\vec{\theta})$
in terms of the complex shear $\gamma(\vec{\theta})$ 
on the whole image plane.

To apply the {\it original} Kaiser and Squires algorithm to real data, 
the following assumptions must be made:\\
(a) The cluster is linear in the sense that $\kappa\ll 1
$ and $|\gamma|\ll 1$ everywhere. \\
(b) Observational data are available
over the entire space. This assumption may be dropped if the cluster is weak 
and small compared to the data field. \\
(c) All background
galaxies have the same {\it effective} redshift; 
i.e., all galaxies have approximately
the same distance ratio $D_{\rm ds}/D_{\rm s}$.

If these assumptions hold, we can perform the direct  
mass reconstruction from the observed distortion field 
$\bar{\epsilon}(\vec{\theta})$.
In the case of weak lensing, the shear $\gamma$ is directly observable:
\begin{equation}
\gamma(\vec{\theta})\approx\bar{\epsilon}(\vec{\theta}).
\end{equation}
In practice, one has to replace the integral in 
Eq. (\ref{KS}) by a sum over a grid $\vec{\theta}_{ij}$ on which the shear is
estimated. 
Using $\Sigma(\vec{\theta})=\Sigma_{\rm cr}\kappa(\vec{\theta})$, 
the surface-mass density   $\Sigma(\vec{\theta})$ of 
the lensing cluster is given by
\begin{eqnarray}
\Sigma(\theta)-\Sigma_0
&\approx& 
\Sigma_{\rm cr}\,\frac{1}{\pi}\int d^2 \theta'\,
\Re[\mbox{$\cal D$}^{*}(\vec\theta-\vec\theta')\,
\bar{\epsilon}(\vec\theta')] \\
&\approx& 
\Sigma_{\rm cr}\,\frac{a^2}{\pi}\sum_{i,j} 
\Re[\mbox{$\cal D$}^*(\vec{\theta}-\vec{\theta}_{ij})
\bar{\epsilon}(\vec{\theta}_{ij})],
\label{linv}
\end{eqnarray} 
where $\Sigma_0$ is a constant surface-mass density which comes from
$\kappa_0$ in Eq. (\ref{KS}),
and $a$ is the separation of the grid points.
In this way, we can reconstruct the cluster-mass distribution from 
the weak-shear
field up to an additive constant $\Sigma_0$.

The crucial result here is that the projected-mass  distribution 
of a cluster 
can be determined only from the observed shear field, that is, only from
information about the shapes of galaxy images. 
However, we must require rather strong assumptions in deriving Eq.
(\ref{linv}).
Here we enumerate several difficulties  which we encounter in performing
the cluster-mass reconstruction 
with the Kaiser and Squires algorithm.\\ 
(1) Application to non-linear clusters:
Since the shear is observable only in the weak-lensing regime, the
inversion algorithm described above cannot be applied to non-linear
clusters. \\
(2) Finite size of the data field:
The integral in the inversion formula  
extends over the entire two-dimensional space, while
real data are available only in a finite region restricted 
by the CCD area. 
Because of the lack of information outside the data field, 
we are forced  to set 
$\gamma=0$ there, which is equivalent for
circularly-symmetric clusters to vanishing total mass within the data field. 
This sharp cut-off of the integration
 leads to an unphysical  negative mass density near the boundary.
We thus suffer from artificial
boundary effects in mass reconstructions owing to the sharp cut-off.\\
%
(3) Redshift distribution of source galaxies:
Cluster-mass reconstructions depend on the assumed redshift of the
background galaxies through the distance ratio $D_{\rm ds}/D_{\rm s}$ in
$\Sigma_{\rm cr}$. Since the redshifts of background galaxies are unknown,
there is a scaling ambiguity in the obtained mass distribution. 
Moreover, if background galaxies are distributed in redshift, the
assumption that the distance ratio $D_{\rm ds}/D_{\rm s}$ is the same for all
background galaxies does not hold, 
especially for high-redshift clusters.\\
(4) Degeneracy of the solution for the cluster-mass inversion:
We have seen that there exists a global 
transformation that leaves observable $\gamma(\vec\theta)$ unchanged in
the weak-lensing limit such that $\kappa(\vec\theta) \to 
\kappa(\vec{\theta})+\kappa_0$, with an arbitrary constant $\kappa_0$;
that is, the surface-mass density is determined only up to an additive
constant from the observed shear field.
We see in the next subsection that there exists a {\it general}
global transformation that leaves the observable unchanged, which we
encounter in all mass-reconstruction schemes based solely on image shapes.

\subsection{Generalization of Kaiser and Squires algorithm for critical
clusters}

We have seen that the original Kaiser and Squires algorithm 
suffers from several difficulties.
In this subsection we concentrate on the difficulties which we face in
applying the Kaiser and Squires algorithm to non-linear clusters, 
where the shear $\gamma$ is no longer a direct observable.

Seitz and Schneider\cite{SeSc95} generalized the Kaiser and Squires algorithm to 
include critical clusters, i.e., clusters which can produce critical 
curves.
This technique, to be discussed in this subsection, 
is based on Eq. (\ref{KS}), and we need the assumptions (b) and (c) 
discussed
in the preceding subsection.

In order to extend the Kaiser and Squires method to the non-linear case,
we express the inversion formula (\ref{KS}) with the reduced shear
$g$. Using $\gamma=g(1-\kappa)$ in Eq. (\ref{KS}), we have 
the integral equation
\begin{equation}
 \kappa(\vec\theta)
  = \frac{1}{\pi}\int d^2 \theta' \,
   \mbox{$\cal D$}^{*}(\vec\theta-\vec\theta')\, g(\vec\theta')
  [1-\kappa(\vec\theta')]. \label{KSwithg}
\end{equation}
Here we have dropped a constant $\kappa_0$ in Eq. (\ref{KSwithg})
because the shear $\gamma$ is not an observable quantity in general.
This equation can be
formally  expressed in an infinite power series as
\begin{eqnarray}
 \kappa(\vec{\theta})
 &=&\mbox{ $\cal G$}-\mbox{ $\cal G$}\circ\!\mbox{ $\cal
  G$}+\mbox{ $\cal G$}\circ\!\mbox{ $\cal
  G$}\!\circ\!\mbox{ $\cal
  G$}-\cdots \nonumber\\
  &=&\sum_{n=1}^{\infty}(-1)^{n-1} \mbox{ $\cal G$}^n ,
\end{eqnarray}
where $\mbox{ $\cal G$}$ is the integral operator
\begin{equation}
\mbox{ $\cal G$}(\vec\theta,\vec\theta')
:=\frac{1}{\pi}\int d^2\theta'\mbox{ $\cal
D$}^*(\vec\theta-\vec\theta')g(\vec\theta')\,\times,
\end{equation}
and $\mbox{ $\cal G$}(\vec\theta,\vec\theta')$ acts on 
a function of $\vec\theta'$.
The Kaiser and Squires method
corresponds to the first-order approximation to this power series in the
weak-lensing limit.

For non-critical clusters, i.e., $\mbox{det}{\cal A}(\vec{\theta})>0$ for
all $\vec{\theta}$, the reduced shear is directly observable:
$g=\bk{\epsilon}_{\epsilon_{\rm
s}}\approx\bar{\epsilon}$. 
For critical
clusters, however, the relation 
between ellipticity and lensing property depends on
the parity of the image (see Eq. (\ref{eg1})), and we cannot 
determine the parity locally. 
To take account of the parity distinction, we write the shear $\gamma$ in
the form
\begin{eqnarray}
\gamma
&=&g\,(1-\kappa) \nonumber \\
&=& H(1-|g|)\,(1-\kappa)\,
\bar{\epsilon}  
+H(|g|-1)\,(1-\kappa)\,
\frac{1}{\bar{\epsilon}^*} \label{gdistinction}.
\end{eqnarray}
Seitz and Schneider\cite{SeSc95} proposed an iterative procedure to 
solve the non-linear inversion equation (\ref{KSwithg}): 
From Eqs. (\ref{KSwithg}) and
(\ref{gdistinction}), we have 
\begin{eqnarray}
 &&\kappa^{(n+1)}(\vec\theta)  \nonumber\\
  &=& \frac{1}{\pi}\int d^2 \theta' \,
H(1-|g^{(n)}(\vec{\theta'})|) \,
[1-\kappa^{(n)}(\vec\theta')]\,
\Re\left[
\mbox{$\cal D$}^{*}(\vec\theta-\vec\theta')\,\bar{\epsilon}(\vec{\theta'})
\right] \nonumber\\
&+&
\frac{1}{\pi}\int d^2 \theta' \,
H(|g^{(n)}(\vec{\theta'})|-1) \,
[1-\kappa^{(n)}(\vec\theta')]\,
\Re\left[
\mbox{$\cal D$}^{*}(\vec\theta-\vec\theta')\,
\frac{1}{\bar{\epsilon}^*(\vec{\theta'})}
\right],
\label{nKS1}
\end{eqnarray}
where $ g^{(n)}=\gamma^{(n)}/(1-\kappa^{(n)})$ 
is the reduced shear in the $n$-th
 step of the iteration, and $\gamma^{(n)}$ is calculated by
\begin{equation}
\gamma^{(n)}=
\frac{1}{\pi}\int d^2 \theta'\,
   \mbox{$\cal D$}(\vec{\theta}-\vec{\theta}')\,\kappa^{(n)}(\vec{\theta}')
\label{nKS2}.
\end{equation}
This iteration is performed by starting with 
$\kappa^{(0)}(\vec{\theta})=\gamma^{(0)}(\vec{\theta})=0$ 
for all $ \vec{\theta}$.
Here we note that the first term on the right-hand side of
Eq. (\ref{nKS1}) 
is the contribution to $\kappa(\vec{\theta})$ from the even-parity 
region, and the second term from the odd-parity region. In practice, 
of course, we use a discretized version of Eqs. (\ref{nKS1}) and (\ref{nKS2}).

In general, the observable quantity is not the
shear $\gamma$ but the reduced shear $g$ (or $1/g*$).
We see that the reduced shear $g$ is invariant under the transformation 
\begin{equation}
\kappa(\vec{\theta})\to\lambda \kappa(\vec{\theta})+(1-\lambda)\,\,\,,\,\,\,
\gamma(\vec{\theta})\to \lambda\gamma(\vec{\theta})\label{invariant}
\label{invtrans}
\end{equation}
with an arbitrary scalar constant $\lambda\neq 0$;\cite{ScSe95} \ 
this transformation is equivalent to scaling 
the Jacobian matrix ${\cal A}(\vec{\theta})$ with $\lambda$:
\begin{equation}
{\cal A}(\vec{\theta})\to \lambda {\cal A}(\vec{\theta})=\lambda
\left(
  \begin{array}{@{\,}cc@{\,}}
 1-\kappa-\gamma_1 & -\gamma_2 \\
 -\gamma_2         & 1-\kappa+\gamma_1
  \end{array}
\right).
\end{equation}
Thus we can determine
$\kappa(\vec{\theta})$ only up to a constant. This is because we use
the information only regarding the image shapes. 
We see that this transformation leaves the critical curves 
$\mbox{det}{\cal A}(\vec{\theta})=0$ invariant. 
This indicates that 
we cannot determine the constant $\lambda$ even if GLAs, 
which are tracers of critical curves, are observed. Further, the curve 
$\kappa(\vec{\theta})=1$, 
on which the gravitational distortions disappear, is left
invariant under the transformation (\ref{invariant}).
We can, however, constrain the value of $\lambda$ 
by requiring that the surface-mass density is
positive everywhere, which yields a lower bound on $\kappa(\vec{\theta})$.
 
A possible method to break this
degeneracy is to employ the information about 
magnification effects.
That is, we employ the fact that the magnification $\mu$ transforms as
\begin{equation}
\mu(\vec{\theta})\to \lambda^{-2}\mu(\vec{\theta})
\end{equation}
under the transformation (\ref{invariant}). 
Since  gravitational magnification changes the size of a galaxy image 
and the number density of
background galaxies, we can in principle 
break the mass degeneracy 
by combining the shear analysis with such magnification 
effects.\cite{SeSc97,Br95,BartN95} 
 
\subsection{Finite field inversion of the cluster mass distribution}

Both the inversion methods described above are based on Eq. (\ref{KS})
which involves the convolution integral of the shear over the entire 
two-dimensional space.
If one tries to apply the inversion formula (\ref{KS}) to real data which are
limited by the finite size of the CCD frame, then 
artificial boundary effects cannot be avoided.

In this subsection we review the finite-field inversion method for the
cluster-mass distribution  derived by 
Seitz and Schneider.\cite{SeSc96} 
This inversion method  has been 
introduced to optimize the observational data, that is, 
to minimize the statistical errors due to noise. 
We assume, for simplicity, 
 that all
background galaxies have about the same distance ratio 
$D_{\rm ds}/D_{\rm s}$ as
in {\S} 4.1 and 4.2. We 
shall discuss the redshift distribution of background galaxies in
{\S} 4.4.

All finite-field methods start from the fact
that the gradient of the convergence $\kappa$ is related to the first
derivatives of the shear $\gamma$:\cite{K95}
\begin{equation}
\vec{\nabla}_{\!\theta}\kappa(\vec{\theta})=
\left(
  \begin{array}{@{\,}c@{\,}}
  \gamma_{1,1} +\gamma_{2,2}\\
  \gamma_{2,1} -\gamma_{1,2}
 \end{array}
\right)\equiv \vec{U}(\vec{\theta}).\label{Kaiser1995}
\end{equation}
In the weak-lensing limit where the shear $\gamma$ is a direct
observable, we can obtain the vector field $\vec{U}(\vec{\theta})$ from
the observed distortion field. However,  
the shear $\gamma$ is not an observable in general.
Inserting $\gamma=g(1-\kappa)$ into Eq. (\ref{Kaiser1995}), we obtain
the relation\cite{K95}
 \begin{equation}
 \vec{\nabla}_{\!\theta}K(\vec{\theta})
=\frac{-1}{1-|g|^2}
\left(
  \begin{array}{@{\,}cc@{\,}}
 1-g_1 & -g_2 \\
 -g_2  & 1+g_1
  \end{array}
\right)
\left(
  \begin{array}{@{\,}c@{\,}}
g_{1,1}+g_{2,2}\\
g_{2,1}-g_{1,2}
\end{array}
\right)\equiv u(\vec{\theta}), \label{Kaiser2}
 \end{equation}
where $K(\vec{\theta})$ is the  scalar field defined by
\begin{equation}
K(\vec{\theta}):=\ln[1-\kappa(\vec{\theta})].
\end{equation}
For a non-critical cluster, we can construct the vector field
$\vec{u}(\vec{\theta})$ in terms of the observable quantity
$g(\vec{\theta})=\bk{\epsilon}_{\epsilon_{\rm s}}
\!(\vec{\theta})\approx\bar{\epsilon}(\vec{\theta})$. 
Hence, the relation (\ref{Kaiser2}) can be used to
reconstruct the surface-mass density from the shear field on a finite
region ${\cal U}$.
Since $\vec{u}(\vec{\theta})$ in Eq. (\ref{Kaiser2}) is a gradient 
field, the differential 
equations (\ref{Kaiser2}) can be solved 
up to an additive constant by integrating $\vec{u}(\vec{\theta})$ from
an arbitrary point $\vec{\theta}_0$ to the point $\vec{\theta}$ along
an arbitrary smooth curve $\vec{l}$:
\begin{equation}
K(\vec{\theta})
=K(\vec{\theta}_0)+\int_{\vec{\theta}_0}^{\vec{\theta}} 
d\vec{l}\cdot
\vec{u}(\vec{l}).\label{line}
\end{equation} 
However, the vector field $\vec{u}(\vec{\theta})$ is obtained from noisy 
data, so that it will contain a rotational component. Therefore, 
$\vec{u}(\vec{\theta})$ is not a gradient field in general; in other
words, mass reconstructions from line integrals of Eq. (\ref{Kaiser1995}) or 
(\ref{Kaiser2}) depend on the choice of the integration path.

Seitz and Schneider\cite{SeSc96} have explicitly taken into account the noise 
component of the vector field $\vec{u}(\vec{\theta})$ obtained from the
observed distortion field by decomposing 
$\vec{u}(\vec{\theta})$ into gradient and  
rotational components:
\begin{equation}
\vec{u}(\vec{\theta})=\vec{\nabla}_{\!\theta}K(\vec{\theta})+
\vec{\nabla}_{\!\theta}\times s(\vec{\theta}),\label{deco}
\end{equation}
where
\begin{equation}
\vec{\nabla}_{\!\theta}\times 
s(\vec{\theta}):=
\left(
  \begin{array}{@{\,}c@{\,}}
 \der s / \der\theta_2  \\
-\der s / \der\theta_1
  \end{array}
\right).\label{rotat}
\end{equation}
Here $s(\vec{\theta})$ is a scalar field which is assumed to account for
the noise part  within the data field ${\cal U}$.  
The decomposition (\ref{deco}) can be determined uniquely by
requiring that the average of $\vec{\nabla}_{\!\theta}\times s(\vec{\theta})$ 
within ${\cal U}$ vanishes, which is reasonable if the rotational
component $\vec{\nabla}_{\!\theta}\times
s(\vec{\theta})$ is due to noise. 
This requirement can be satisfied if $s(\vec{\theta})$ satisfies the
condition
\begin{equation}
s(\vec{\theta })={\rm const} \ 
\mbox{on the boundary $\der{\cal U}$ of $\cal U$}.\label{noise s}
\end{equation}
(This is not a necessary but a sufficient condition.)
If the condition (\ref{noise s}) is satisfied, the scalar field
$K(\vec{\theta})$ can be obtained from the
vector field (\ref{deco}):
\begin{equation}
K(\vec{\theta})-\bar{K}=\int_{\cal U}d^2\theta'\,\vec{H}(\vec{\theta}',
\vec{\theta})\cdot\vec{u}(\vec{\theta'}) \label{KernelH}
\end{equation}
with 
\begin{equation}
\vec{H}(\vec{\theta'},\vec{\theta}):=-\vec{\nabla}_{\!\theta'} {\cal L}
(\vec{\theta'},
\vec{\theta}).
\end{equation}
Here $\bar{K}$ is a constant which represents the 
average of $K(\vec{\theta})$ within the data
field ${\cal U}$, and ${\cal L}(\vec{\theta'},\vec{\theta})$ is 
the solution of the Neumann boundary problem
\begin{eqnarray}
&&\triangle_{\theta'}
{\cal L}(\vec{\theta'},\vec{\theta})=\delta^2\!(\vec{\theta}-\vec{\theta}')
-\frac{1}{A}\,\,\,\,\,\,\forall\vec{\theta} \in {\cal U} \\
&&\vec{n}(\vec{\theta}')\cdot\vec{\nabla}_{\!\theta'}
{\cal L}(\vec{\theta'},\vec{\theta})=0\,\,\,\,\,\,
\forall\vec{\theta'}\in {\der\cal U},
\end{eqnarray}
where $A$ is the area of the data field ${\cal U}$, and
$\vec{n}$ is the unit vector orthogonal to the boundary $\der{\cal U}$;
${\cal L}(\vec{\theta'},\vec{\theta})$ is solved uniquely up to an
additive constant, so that the kernel
$\vec{H}(\vec{\theta'},\vec{\theta})$ is determined uniquely for a given 
boundary $\der{\cal U}$. 
The solution for two special geometries, a circle and a rectangle, 
is derived in Ref.~\citen{SeSc96}. 
The kernel $\vec{H}(\vec{\theta'},\vec{\theta})$ is called 
{\it noise filtering} because it filters out the rotational noise 
component. 
We note that we can replace
$(K,\vec{u})$ in Eq. (\ref{KernelH}) with $(\kappa,\vec{U})$,
because $\kappa$ and $\vec{U}$
are related to each other through the same relation
($\vec{\nabla}_{\!\theta}\kappa=\vec{U}$) as 
that relating  $K$ and $\vec{u}$ 
($\vec{\nabla}_{\!\theta}K=\vec{u}$); that is, 
\begin{equation}
\kappa(\vec{\theta})-\bar{\kappa}=
\int_{\cal U}d^2\theta'\,\vec{H}(\vec{\theta}',
\vec{\theta})\cdot\vec{U}(\vec{\theta'}), \label{KernelH2}
\end{equation}
where $\bar{\kappa}$ is the average of $\kappa(\vec{\theta})$ within the 
data field ${\cal U}$.
We see from Eqs. (\ref{KernelH}) and (\ref{line}) 
that $K(\vec{\theta})=\ln[1-\kappa(\vec{\theta})]$ 
can be obtained up to an additive constant $\bar{K}$; i.e., 
$\kappa(\vec{\theta})$ is determined only up to the global transformation  
\begin{equation}
\kappa(\vec{\theta})\to\lambda\kappa(\vec{\theta})+(1-\lambda)
\end{equation}
with an arbitrary constant $\lambda\neq 0$.

Several authors have developed inversion techniques 
+that require data only on a finite
field.\cite{K95,Bart95,Sc95,SeSc96,SqK96} \   
Seitz and Schneider\cite{SeSc96}
have shown that there exist an infinite number of finite-field
formulae, which are mathematically equivalent but different in their
dependence on the noise due to the
discreteness of data or the intrinsic ellipticity of the background
galaxies. 
They also analyzed quantitatively 
the power spectra of the error fields for the
mass maps reconstructed with synthetic data using
different inversion methods: the non-linear
version of the Kaiser and Squires inversion equation (\ref{nKS1}), 
the finite-field inversion equation with the noise-filtering kernel 
and two other finite-field inversion equations.
The main results obtained there are (i)
the inversion formula (\ref{KernelH}) 
performs better than the  other two finite-field
formulae, (ii) the noise for the finite-inversion (\ref{KernelH}) is
more uniform than that for the other inversions over the data field and
(iii) concerning the small-scale error, 
the modified Kaiser and Squires inversion (\ref{nKS1}) 
works best of all the inversion tested there,
while it suffers from artificial boundary effects. 
They concluded from
their results that
the finite-field inversion with the noise-filtering kernel 
is the most accurate of 
all the
inversion formulae used in their analysis. 
Lombardi and Bertin\cite{LB98} have 
obtained expressions for the errors on the mass maps reconstructed 
with various inversion formulae and shown
that a rotation-free kernel, such as the noise-filtering kernel,
should be used in order to optimize the reconstruction procedure.
The analytical results obtained by Lombardi and Bertin 
confirm the conclusion of Seitz and Schneider.\cite{SeSc96}

\subsection{The General cluster mass reconstruction algorithm}

Up to this point we have only dealt with the case where
 all background 
galaxies have the same effective redshift, i.e., 
all background galaxies have about the same distance ratio 
$D_{\rm ds}/D_{\rm s}$. 
This approximation
is fairly good if the cluster redshift is relatively low 
 and 
almost all galaxies have high redshifts. 
For high-redshift
clusters, however, the redshift distribution of the
source galaxies has to be taken into account in the inversion
procedure explicitly.
We review in this subsection the 
mass-reconstruction scheme developed 
by Seitz and Schneider,\cite{SeSc97} \ 
which includes the case where background galaxies are distributed in
redshift.
Here we fix the lens redshift $z_{\rm d}$ and express 
lensing properties explicitly as functions of the source redshift $z$ 
(e.g., $\kappa(\vec{\theta})\equiv\kappa(\vec{\theta},z)$).

If we define the {\it relative lensing strength} $w(z)$ for a source 
with redshift $z$ by
\begin{equation}
w(z):=
H(z-z_{\rm d})\,
\frac{\Sigma_{{\rm cr},\infty}}{\Sigma_{\rm cr}(z)}
\end{equation}
with 
\begin{equation}
\Sigma_{{\rm cr},\infty}:=\lim_{z\to\infty}\Sigma_{\rm cr}(z),
\end{equation}
then the convergence and the shear for a source at redshift $z$ 
can be expressed as   
\begin{equation}
\kappa(\vec{\theta},z)
=H(z-z_d)
\frac{\Sigma(\vec{\theta})}
      {\Sigma_{\rm cr}(z)}
=w(z)
 \frac{\Sigma(\vec{\theta})}
      {\Sigma_{{\rm cr},\infty}}\equiv
w(z)\,\kappa_{\infty}(\vec{\theta})\label{kapz}
\end{equation}
and
\begin{equation}
\gamma(\vec{\theta},z)\equiv w(z)\,\gamma_{\infty}(\vec{\theta}),\label{gamz}
\end{equation}
respectively. 
Note that
$\kappa(\vec{\theta},z)=\gamma(\vec{\theta},z)=0$ for a source with
redshift $z\leq z_{\rm d}$.
The function $w(z)$ can be calculated for a given background cosmology;
for an Einstein-de Sitter universe ($\Omega_0=1$), $w(z)$ is given by
\begin{equation}
w(z)=H(z-z_{\rm d})\, \frac{\sqrt{1+z}-\sqrt{1+z_{\rm d}}}{\sqrt{1+z}-1}, 
\label{wzEde} 
\end{equation}
and $\Sigma_{{\rm cr},\infty}$ is calculated to be
\begin{eqnarray}
\Sigma_{{\rm cr},\infty}
&=&\frac{c H_0}{8\pi G}\frac{(1+z_{\rm d})^2}{\sqrt{1+z_{\rm d}}-1}\nonumber\\
&=&1.39 \times 10^{14} \frac{(1+z_{\rm d})^2}{\sqrt{1+z_{\rm d}}-1}  
h_{50}\;\;M_{\odot} \ {\rm Mpc}^{-2}. \label{sigmacrinf}
\end{eqnarray}
From Eqs.  (\ref{mag}), (\ref{kapz}) and (\ref{gamz}), the magnification
$\mu(\vec{\theta},z)$ 
of an image at angular position $\vec{\theta}$ and redshift $z$ is given 
by
\begin{eqnarray}
\mu(\vec{\theta},z)&=&
\frac{1}{
|\mbox{det}{\cal A}(\vec{\theta},z)|}\nonumber\\
&=&
\frac{1}{\left|
[1-w(z)\,\kappa_{\infty}(\vec{\theta})]^2-
w^2(z)\,|\gamma_{\infty}(\vec{\theta})|^2\right|}.
\end{eqnarray}
If $\mbox{det}{\cal
A}(\vec{\theta},\infty)=
[1-\kappa_{\infty}(\vec{\theta})]^2-|\gamma_{\infty}(\vec{\theta})|^2
>0$ for all $ \vec{\theta}$, then the cluster is non-critical for all 
source redshifts.

As we have seen in {\S} 3, only $g$ or $1/g^*$ is accessible to the
observable quantity $\epsilon$.
For a fixed source redshift $z$, 
the expectation value of
$\epsilon^n(\vec{\theta})$ is 
\begin{equation}
\bk{\epsilon^n}_{\epsilon_{\rm s }}(\vec{\theta},z)
=
\left\{
\begin{array}{@{\,}lll}
 g^n(\vec{\theta},z)         &\mbox{for} & |g(\vec{\theta},z)| \leq 1,  \\
 1/ {g^*}^n(\vec{\theta},z)  &\mbox{for} & |g(\vec{\theta},z)|    > 1
\end{array}
\right. 
\end{equation}
(see Eq. (\ref{eg3})), where 
\begin{equation}
g(\vec{\theta},z)=\frac{\gamma(\vec{\theta},z)}{1-\kappa(\vec{\theta},z)}
=\frac{w(z)\,\gamma_{\infty}(\vec{\theta})}
{1-w(z)\,\kappa_{\infty}(\vec{\theta})} .
\end{equation}
We then consider the case where source galaxies are distributed in
redshift. 
Let $p_z(z)\,dz$ be
the probability that the source redshift is within $dz$ around
$z$; this is a simple approximation, 
because the gravitational magnification 
changes the observed number density of galaxies, so that the probability 
distribution of the observed source redshift depends on the
magnification $\mu(\vec{\theta},z)$.\cite{Br95} \ 
This
simplification is justified
if the overall magnification is found to be small, or if the dependence
of the galaxy 
redshift distribution on the flux is weak. 

If $p_z(z)$ is assumed to be known, 
the expectation value of $\epsilon^n$ can be expressed by
\begin{eqnarray}
\bk{\epsilon^n}_{\epsilon_{\rm s},z}
&:=&
\int_0^{\infty}\!dz\,p_z(z)\!\int\!d^2\epsilon_{\rm s}\,p_{\epsilon_{\rm s}}
\epsilon^n(\epsilon_{\rm s},g(z)) \nonumber\\
%
&=&\int_{\mbox{det}{\cal A}(z)\geq 0} 
\! dz\,p_z(z)\,g^n(z)+
\int_{\mbox{det}{\cal A}(z)< 0}
\!dz\,p_z(z)\,\left(\frac{1}{g^*(z)}\right)^n \nonumber\\
&=&\gamma_{\infty}^n\int_{\mbox{det}{\cal A}(w)\geq 0}
\! dw\,p_w(w)\,\left(\frac{w}{1-\kappa_{\infty} w}\right)^n  \nonumber\\
&&+
\frac{\gamma_{\infty}^n}{|\gamma_{\infty}|^{2n}}
\int_{\mbox{det}{\cal A}(w)< 0}
\! dw\,p_w(w)\,\left(\frac{1-\kappa_{\infty} w}{w}\right)^n\nonumber\\
&\equiv&
\gamma_{\infty}^n\left[
 X_n(\kappa_{\infty},\gamma_{\infty})+\frac{1}{|\gamma_{\infty}|^{2n}}
Y_n(\kappa_{\infty},\gamma_{\infty})
\right], \label{expectation}
\end{eqnarray}
where $p_w(w)$ is the probability distribution function of $w$, which is 
given by $p_w(w)\,dw=p_z\,dz$. 
Here the functions $X_n$ and $Y_n$ depend on 
$\kappa_{\infty}$ and $\gamma_{\infty}$ by
\begin{eqnarray}
X_n(\kappa_{\infty},\gamma_{\infty})
&=&\left(\int_0^{\min(1,1/(\kappa_{\infty}+|\gamma_{\infty}|))}
  +\int_{\frac{1}{\max(1,\kappa_{\infty}-|\gamma_{\infty}|)}}^1
\right)\nonumber\\
&&\times dw\,p_w(w)\left(\frac{w}{1-\kappa_{\infty} w}\right)^n,\\
Y_n(\kappa_{\infty},\gamma_{\infty})
&=&\int_{\min(1,1/(\kappa_{\infty}+|\gamma_{\infty}|))}
^{\frac{1}{\max(1,\kappa_{\infty}-|\gamma_{\infty}|)}}\!dw\,p_w(w)\,
\left(\frac{1-\kappa_{\infty} w}{w}
\right)^n.
\end{eqnarray}
Now we have the local relation between the
expectation value of $\epsilon$ and the lensing properties,
\begin{equation}
\gamma_{\infty}(\vec{\theta})
=\bk{\epsilon}_{\epsilon_{\rm s},z}(\vec{\theta})\left[
X_1(\kappa_{\infty}(\vec{\theta}),\gamma_{\infty}(\vec{\theta}))+
\frac{1}{|\gamma_{\infty}(\vec{\theta})|^2}Y_1(\kappa_{\infty}(\vec{\theta}), 
\gamma_{\infty}(\vec{\theta}))
\right]^{-1}. \label{evsl}
\end{equation}
In a practical application, the redshift-averaged ellipticity
$\bk{\epsilon}_{\epsilon_{\rm s},z}(\vec{\theta})$ 
must be replaced by the observed ellipticity $\bar{\epsilon}(\vec{\theta})$
defined by Eq. (\ref{epsbar}).
In the weak-lensing limit (i.e., $\kappa_{\infty} \ll 1$ and 
$|\gamma_{\infty}|\ll 1$), we have
\begin{equation}
X_n\,(\kappa_{\infty},\gamma_{\infty})
\approx \bk{w^n} \ ; \ 
Y_n(\kappa_{\infty},\gamma_{\infty})=0
\end{equation} 
with
\begin{equation}
\bk{w^n}:=\int_0^1 \!dw\,p_w(w)\,w^n,
\end{equation}
and therefore the shear $\gamma_{\infty}$ is observable: 
\begin{equation}
\gamma_{\infty}(\vec{\theta})\approx
\bar{\epsilon}(\vec{\theta})/\bk{w}. \label{zlinear}
\end{equation}
That is, in the weak-lensing limit, 
the situation is the same as in the case where 
all sources have the same redshift $z$ such that $w(z)=\bk{w}$.

In {\it general} cluster-mass reconstructions, the finite-field
inversion formula (\ref{KernelH2}) can be used:\cite{SeSc97} 
\begin{equation}
\kappa_{\infty}(\vec{\theta})-\bar{\kappa}_{\infty}=
\int_{\cal U}d^2\theta'\,\vec{H}(\vec{\theta}',
\vec{\theta})\cdot\vec{U_{\infty}}(\vec{\theta'}) \label{ff}
\end{equation}
with
\begin{equation}
\vec{U}_{\infty}(\vec{\theta}):=
\left(
  \begin{array}{@{\,}c@{\,}}
(\gamma_{\infty})_{1,1} +(\gamma_{\infty})_{2,2}\\
(\gamma_{\infty})_{2,1} -(\gamma_{\infty})_{1,2}
\end{array}
\right). \label{Uinf}
\end{equation}
Here $\bar{\kappa}_{\infty}$ is the (unknown) average of
$\kappa_{\infty}(\vec{\theta})$ within the data field ${\cal U}$.
Inserting Eq. (\ref{evsl}) in Eq. (\ref{ff}), 
we have the integral equation for $\kappa_{\infty}(\vec{\theta})$, which 
can be solved iteratively. Starting the iteration with
$\kappa^{(0)}_{\infty}(\vec{\theta})=\gamma^{(0)}_{\infty}(\vec{\theta})=0$ 
for all $\vec{\theta}$, we calculate $\kappa_{\infty}^{(n)}$ and
$\gamma_{\infty}^{(n)}$ for $n\geq 1$ by
\begin{eqnarray}
&&\kappa^{(n+1)}_{\infty}(\vec{\theta})-\bar{\kappa}_{\infty}=
\int_{\cal U}d^2\theta'\,\vec{H}(\vec{\theta}',
\vec{\theta})\cdot\vec{U}_{\infty}^{(n+1)}(\vec{\theta'}),\label{kapn1}\\
&&\gamma_{\infty}^{(n+1)}
=\bar{\epsilon}\left[
X_1(\kappa^{(n)}_{\infty},\gamma^{(n)}_{\infty})+
\frac{1}{\left|\gamma^{(n)}_{\infty}\right|^2}
Y_1(\kappa^{(n)}_{\infty}, 
\gamma^{(n)}_{\infty})
\right]^{-1}.
\end{eqnarray}
Here the vector field $\vec{U}_{\infty}^{(n+1)}(\vec{\theta})$ is
calculated from Eq. (\ref{Uinf}) with
$\gamma_{\infty}^{(n+1)}(\vec{\theta})$.

 It can be seen from Eq. (\ref{kapn1}) that
this iterative procedure  still contains the undetermined constant $\bar{
\kappa}_{\infty}$. Hence, this inversion algorithm 
involves a global transformation that leaves the observable
invariant. 
In addition, the dependence of the resulting mass distribution on the
constant $\bar{\kappa}_{\infty}$ is highly non-linear, so that this
transformation cannot be determined analytically. 
In the case of weak lensing ($\kappa_{\infty}\ll 1$ and
$\gamma_{\infty}\ll1$), we have seen that the mass reconstruction
depends only on $\bk{w}$, so that the global invariance transformation
is equivalent to adding a constant. 
If the cluster is non-critical for
all sources (i.e., $\mbox{det}{\cal A}(\infty,\vec{\theta})>0$ for all 
$\vec{\theta})$, then $Y_n$ vanishes and $X_n$ depends only on
$\kappa_{\infty}$, yielding a somewhat simple relation:
\begin{equation}
\bk{\epsilon^n}_{\!\epsilon_{\rm s},z}=\gamma^n_{\infty}\,X_n(\kappa_{\infty}).
\end{equation}
On the other hand, 
an approximation for the function $X_1(\kappa_{\infty})$ is
\begin{equation}
X_1(\kappa_{\infty})\approx
\frac{\bk{w}}{1-\kappa_{\infty}\frac{\bk{w^2}}{\bk{w}}}, 
\end{equation}
which is quite accurate for non-linear but not very strong
clusters.\cite{SeSc97} 
In this case,  the global invariance transformation 
becomes
\begin{equation}
\kappa_{\infty}(\vec{\theta})\to\lambda\kappa_{\infty}(\vec{\theta})
+\frac{(1-\lambda)\bk{w}}{\bk{w^2}} \label{approxinv}
\end{equation}
with an arbitrary constant $\lambda\neq 0$.

\subsection{Cluster mass estimates from shear fields}

In the previous subsections we have described the cluster-mass
reconstruction methods based on the shear analysis. One of
the goals of such studies is to obtain the total mass of a cluster 
within a given region.
In estimating the cluster mass inside a given aperture, one method to 
obtain the aperture mass is to use the reconstructed two-dimensional 
mass distribution. 
However, the resulting mass map from the shear field  will be noisy
because of the discreteness of the galaxy images, intrinsic
source ellipticities, etc. 
In addition, it is difficult to estimate an error for the 
local convergence
since the errors of the reconstructed convergence 
at different points will be strongly correlated.
In this subsection we describe a method to infer the projected cluster
mass  
inside circular apertures from weak lensing
without reconstructing the entire
mass distribution.

First, we define a polar coordinate system ($\vartheta, \varphi$) 
centered on a point $\vec{\theta}_0$ such that
\begin{equation}
\vec{\theta}(\vartheta,\varphi)=
\left(
  \begin{array}{@{\,}c@{\,}}
 \vartheta\cos\varphi \\
 \vartheta\sin\varphi
  \end{array}
\right)+\vec{\theta}_0.
\end{equation} 
Then the convergence
averaged within a circle of radius $\vartheta$ around $\vec{\theta}_0$
is given by
\begin{equation}
\bar{\kappa}(\vartheta)
:=
\frac{1}{\pi\vartheta^2}\int_0^\vartheta d\vartheta'\,
\vartheta'
\oint d\varphi'\,
\kappa(\vec{\theta}(\vartheta',\varphi')).
\end{equation}
Using the two-dimensional version of Gauss's theorem, 
this quantity can be transformed into 
\begin{equation}
\bar{\kappa}(\vartheta)
=\bk{\kappa}(\vartheta)+
\bk{\gamma_{\rm t}}(\vartheta;\vec{\theta}_0), 
\label{kapvsgamt}
\end{equation}
where 
the angular brackets denote the average over a circle of radius $\vartheta$,
e.g.,
\begin{equation}
\bk{\gamma_{\rm t}}(\vartheta;\vec{\theta}_0)
:=
\oint \frac{d\varphi}{2\pi}\,
\gamma_{\rm t}(\vec{\theta}(\vartheta,\varphi);\vec{\theta}_0)
\end{equation}
with the tangential shear $\gamma_{\rm t}(\vec{\theta};\vec{\theta}_0)$
relative to the origin $\vec{\theta}_0$ of the coordinate system.

On the other hand, $\bar{\kappa}(\vartheta)$ 
can be expressed as
\begin{equation}
\bar{\kappa}(\vartheta)
=\frac{2}{\vartheta^2}\int_0^\vartheta d\vartheta'\,
\vartheta'\bk{\kappa}(\vartheta'). \label{kappaaverage}
\end{equation}
From  Eqs. (\ref{kapvsgamt}) and (\ref{kappaaverage}), we have 
\begin{equation}
\frac{d\bar{\kappa}(\vartheta)}{d\ln\vartheta}
=-2\bk{\gamma_{\rm t}}(\vartheta;\vec{\theta}_0) \label{kvsgdif}.
\end{equation} 
Integrating Eq. (\ref{kvsgdif})
between radii $\vartheta$ and
$\vartheta'$ ($ > \vartheta$), 
we obtain the so-called {\it $\zeta$-statistic},\cite{F94}
\begin{equation}
\zeta(\vartheta,\vartheta')
:=\bar{\kappa}(\vartheta)-
  \bar{\kappa}(\vartheta,\vartheta')
=\frac{2\vartheta'^2}{\vartheta'^2-\vartheta^2}
\int_{\vartheta}^{\vartheta'} \frac{d\vartheta''}{\vartheta''} 
\,\bk{\gamma_{\rm t}}(\vartheta'';\vec{\theta}_0),\label{zeta}
\end{equation}
where $\bar{\kappa}(\vartheta,\vartheta')$
 is the average of the convergence
within an annulus between $\vartheta$ and $\vartheta'$:
\begin{equation}
\bar{\kappa}(\vartheta,\vartheta')
:=\frac{1}{\pi(\vartheta'^2-\vartheta^2)}
\int_{\vartheta}^{\vartheta'}d\vartheta''\,\vartheta''
\bk{\kappa}(\vartheta'').
\end{equation}
In the weak-lensing limit (i.e., $\kappa \ll 1$ and $|\gamma|\ll 1$),
the $\zeta$-statistic can be determined uniquely from the shear field,
because the invariance transformation, which corresponds to an additive
constant to $\kappa$, cancels out on the right-hand side of Eq. (\ref{zeta}). 

Since
$\bar{\kappa}(\vartheta,\vartheta')$ has a non-negative value, 
$\zeta(\theta,\theta')$ yields a lower limit on 
$\bar{\kappa}(\vartheta)$;
that is, the quantity 
$
\pi(D_{\rm d}\vartheta)^2 \Sigma_{\rm cr}\zeta(\vartheta,\vartheta')
$
yields a lower limit on the lensing mass 
inside a circular aperture of radius $\vartheta$.
Thus we can infer the lensing mass inside a circular boundary
from the data outside the boundary in the weak-lensing limit.
The $\zeta$-statistic is useful in estimating the total cluster mass 
within the data field.
  
We note that the shape of an aperture need not be restricted to a 
circle. Aperture masses for arbitrary aperture shapes are 
dealt with in Ref.~\citen{ScB97}.


\section{Observational Studies of Weak Lensing}

This section reviews the observational studies of galaxy clusters
based on the weak-lensing analysis. In {\S} 5.1 we summarize some
observational results from weak lensing. 
In {\S} 5.2 we 
discuss the prospects of 
weak-lensing analysis on clusters
with the upcoming $8.3$ m $\it Subaru$ telescope.

\subsection{Observational results}

The non-parametric methods for cluster-mass reconstructions 
by means of weak lensing
have been widely used in investigating cluster properties 
in recent years.
The pioneering work in this field 
was carried out by
Kaiser and
Squires,\cite{KS93} \
 who derived an exact inversion equation for the two-dimensional
mass density of the deflector 
in terms of the tidal component of the gravitational field, namely the shear
which is observable in the weak-lensing regime.
The direct mass-reconstruction method of Kaiser and Squires 
and its variants 
were
then applied to real clusters. 
In practice, 
these mass-reconstruction methods were mainly used to obtain the  
morphology of the mass distribution in a cluster. 
On the other hand, 
the $\zeta$-statistic was often used in the cluster-mass estimations,
as mentioned in {\S} 4.5.
We summarize in Table I some results of 
the weak-lensing analysis on clusters.

Fahlman et al.\cite{F94} 
first applied the Kaiser and Squires method to the
X-ray luminous cluster MS 1224+20 at redshift 0.33. 
The shear field for this cluster was measured from the observational data 
obtained with the $3.6$ m $\it{Canada}$-$\it{France}$-$\it{Hawaii}$
$\it{Telescope}$ (CFHT), and the two-dimensional mass map was derived 
from the shear field.
The location of the main peak of the resulting mass map 
is in good agreement with that of the optical and X-ray centroid.
The mass estimates were performed using the $\zeta$-statistic
(\ref{zeta}) within a circular aperture of radius $2.\am 76$ (corresponding 
to a physical radius of $0.96\;h_{\rm 50}^{-1}$ Mpc for an 
Einstein-de Sitter universe) centered on the mass peak. 
They obtained a lower bound on the aperture mass interior to radius 
$0.96\; h_{\rm 50}^{-1}$ Mpc to be $\simeq 7.0\times 10^{14}\; h_{50}^{-1} 
M_{\odot}$, which
corresponds to a mass-to-light ratio of $M/L\sim 400\;h_{50}$ in solar units.
The mass estimate for this cluster from weak-lensing analysis 
is about three times larger than
that from virial analysis.

Smail et al.\cite{Sm94} analyzed three clusters which span a wide
redshift range, $z_{\rm d}\in [0.26,0.89]$.
Two of these clusters (Cl 1455+22, $z_{\rm d}=0.26$; Cl
0016+16, $z_{\rm d}=0.55$) were selected 
for their high X-ray luminosities, and  the other 
(Cl 1603+43, $z_{\rm d}=0.89$) was optically discovered 
and has a low X-ray luminosity. 
They detected clear lensing signals in the two lower-redshift
clusters, while no significant lensing signal 
was detected in the highest-redshift cluster Cl 1603+43. 
%
They concluded from their results
that
the absence of any lensing signal for their highest-redshift cluster
is ascribed to the absence of a high-redshift population of background
galaxies with $I\le 25$.
On the other hand, Luppino and Kaiser\cite{LK96} 
argued that the failure of Smail et al. to 
detect weak lensing in Cl 1603+43 is simply due to the fact that 
this cluster is not massive enough to produce a detectable shear signal.
The detection of significant shear signals in high-redshift clusters
(MS 1054-03, $z_{\rm d}=0.83$;\cite{LK96} MS 1137+66, $
z_{\rm d}=0.783$;\cite{Cl98} RXJ 1716+67,
$z_{\rm d}=0.813$\cite{Cl98})  
indicates that the great part of background galaxies
are at redshifts of the order of unity, 
supporting the argument by Luppino and Kaiser.

Seitz et al.\cite{Se96} applied a direct mass-reconstruction method to 
the distant cluster Cl 0939+47 at redshift $z_{\rm d}=0.41$, observed
with the {\it Wide Field Planetary Camera 2} (WFPC2) 
on the ${\it Hubble}$ ${\it
Space}$ ${\it Telescope}$ ({\sl HST}). 
Owing to the high redshift of this
cluster and the small field-of-view of the WFPC2, 
the mass-reconstruction scheme described in {\S} 4.4
---which takes into account 
the redshift distribution of the background galaxies
and is based on
the finite-field inversion equation---
was used. 
Assuming a particular form of the parameterized distribution function 
of the source redshift, they reconstructed 
two-dimensional mass maps for several sets of the parameters.
They found a strong correlation between the reconstructed mass
map and the light distribution of the bright cluster galaxies.
The main mass and light maximum agree well with a maximum in the X-ray
image from the {\sl ROSAT}/PSPC observation.\cite{SW96} \ 
They also found a magnification effect on the observed number density
of background galaxies, by which 
they removed the mass degeneracy.

Direct mass-reconstruction methods have also  been applied to the QSO
fields. From the CFHT observation,
Fisher et al.\cite{Fis97} measured a weak gravitational shear
induced by the cluster at a redshift of 0.355 
in the field of the double QSO 0957+561, which is 
a multiply-lensed system with a well-studied time delay.
They reconstructed a two-dimensional mass map in this field from the
measured shear field in order to construct a detailed model for the lens 
system which consists of a primary lensing galaxy and the cluster
containing this galaxy. The resulting mass distribution is consistent
with a spherical cluster 
possessing a
mass peak slightly offset from the
position of the primary lensing galaxy. The obtained projected mass 
within a circular aperture of radius $1\; h_{\rm 50}^{-1}$ 
Mpc centered on the mass peak is
$3.9\pm 1.2 \times 10^{14}\; h_{\rm 50}^{-1} M_{\odot}$. 

Several authors have compared cluster-mass estimates from weak lensing
and those from other methods. Detailed quantitative comparisons of
cluster-mass 
estimates on scales $\sim 1$ Mpc 
show that mass estimates from weak-lensing analysis are consistent with those
from X-ray analysis based on the hydrostatic equilibrium of the ICM with
the gravitational potential 
(Abell 2218;\cite{Sq96a} \ 
 Abell 2390;\cite{Sq96b} \ 
 Abell 2163\cite{Sq97,A98}). 



\subsection{Observation with the Subaru telescope}

The weak-lensing analysis of galaxy clusters requires quite a large number
density ($\sim 50$ arcmin$^{-2}$)
of galaxy images with sufficient detail and accuracy in order to obtain a
reliable shear measurement.
Hence, cluster-mass reconstructions based on the weak-shear field
rely on the power of telescopes and instruments to be used. 
Moreover, for ground-based observations, 
the effect of atmospheric seeing---the circularization of galaxy
images---will reduce the lensing strength considerably
since most of the background galaxies are quite small and faint.
Hence, an excellent seeing condition (sub-arcsec) 
as well as a high angular
resolution is
required for the ground-based weak-lensing analysis.
 
On the other hand, the space-based deep {\sl HST}/WFPC2 observation 
provides us with a
large galaxy-number density without the seeing effect.
Hence, it is a natural consequence that {\sl HST}/WFPC2 
 observations have been
frequently used in weak-lensing analysis 
in recent years.\cite{Se96,Sm97,H98} \ 
However, in contrast to these advantages, 
the {\sl HST}/WFPC2 observations have a
fatal weakness from the theoretical point of view.
As mentioned above, a mass reconstruction based only on the shear
field suffers from mass degeneracy. If the cluster-mass distribution 
extends beyond the observed data field, we cannot uniquely determine the
solution for the cluster-mass inversion from the shear field. In this
case, we obtain
only a lower limit on the cluster mass in the observed field.
Moreover,  since the dependence 
of mass reconstructions on the free parameter $\bar{\kappa}_{\infty}$ 
(see {\S} 4.3) is highly non-linear 
for a high-redshift and super-critical cluster, 
even the resulting morphology of the
cluster could be uncertain, depending on the choice of the parameter.
In addition,
the irregular shape of the WFPC2 frame will also make a mass
reconstruction difficult to realize.  
Hence, both the requirements of a {\it large galaxy-number density} 
and a {\it wide field-of-view}  are needed for  a reliable mass
reconstruction.

The  {\it Subaru telescope} 
is among such telescopes which ideally satisfy both the requirements. 
Subaru is the Japanese 8.3 m  optical-infrared  ground-based
telescope at the summit of Mauna Kea, Hawaii.  
Among the observational instruments for Subaru, 
Suprime-Cam (Subaru Prime Focus
Camera) is the most suitable for weak-lensing analysis.
Suprime-Cam at the F/2.3 prime focus has a wide-field imaging, 
covering a $30'\times 24'$ field-of-view
with a $0.\as 2$/pixel resolution.\cite{SupCam} \ 
This corresponds to a physical scale of about $7.6 \times 6.1
h_{50}^{-2}$ Mpc$^2$  
at a moderate redshift of 0.2 for an Einstein-De Sitter universe. 
Weak-lensing analysis using 
the Subaru telescope with Suprime-Cam can probe
the 
mass distribution on large scales ($\sim 10$ Mpc) where the formation of
gravitational
structures is still in progress.  
Such studies can provide unique, invaluable information on the
distribution of dark matter and $\Omega_0$.
For example, the super cluster MS0302+17 containing three clusters (Cl 
0303+1706, $z_{\rm d}=0.418$; MS 0302+1659, $z_{\rm d}=0.426$; MS
0302+1717, $z_{\rm d}=0.425$   ) is of interest to investigate 
such problems.\cite{K98} 


\clearpage

\begin{table}
 \begin{center}
\caption{
Summary of observational studies of clusters from weak lensing.
}
\begin{tabular}{lllllll}
\hline
\hline
  Cluster & $z_{\rm{d}}^{\,\,\makebox{\scriptsize a)}}$ & 
  r$^{\,\,\makebox{\scriptsize b)}}$         & 
  $M(<r)^{\,\,\makebox{\scriptsize c)}}$         & 
  $ M/L^{\,\,\makebox{\scriptsize d)}}$      & 
  Telescope$^{\,\,\makebox{\scriptsize e)}}$ & 
  Ref.$^{\,\,\makebox{\scriptsize f)}}$ \\
   &    & $( h_{50}^{-1})$ & $(h_{50}^{-1})$ & $(h_{50})$& & \\
  \hline
Abell 1689 & 0.184 &3 &89   & 200$\pm$30 &CTIO$^{{\rm \  i})}$&
 ~\citen{TF95,WF97} \\
Abell 2163 & 0.208 &0.90& 13$\pm$7 & 150$\pm$ 50 &CFHT$^{{\rm \ ii})}$ &
 ~\citen{Sq97,WF97}\\
Abell 2218 & 0.175 &0.8& 7.8$\pm$1.4 & 220$\pm$ 40 &CFHT &
 ~\citen{Sq96a} \\
      &       &0.4& 2.10$\pm$0.38& 155 &{\sl HST}$^{{\rm \ iii})}$ &   
~\citen{Sm97}\\   
Abell 2390 & 0.231 &  1.15&19.5$\pm$6.5& 160$\pm$45&CFHT&
 ~\citen{Sq96b,WF97} \\
AC 118  & 0.308  &0.4 &3.70$\pm$0.64 & 185 &{\sl HST} & ~\citen{Sm97}\\ 
($=$Abell 2744) & & & & & &\\
Cl 0016+16& 0.546 & 0.6& 8.5& 370 &WHT           & ~\citen{Sm95}\\
      &      & 0.4& 3.74$\pm$1.28& 90& {\sl HST} & ~\citen{Sm97}\\
Cl 0024+16& 0.39 &0.4 &2.78$\pm$0.74& 75 &{\sl HST}  & ~\citen{Sm97}\\
Cl 0054-27& 0.56 &0.4 &3.42$\pm$1.28 & 200 &{\sl HST}       & ~\citen{Sm97}\\
Cl 0303+17& 0.42 &0.4 &0.44$\pm$0.90 & 40 &{\sl HST}        & ~\citen{Sm97}\\
Cl 0412-65& 0.51 &0.4 &0.50$\pm$0.82 & 35 &{\sl HST}       & ~\citen{Sm97}\\
Cl 0939+47& 0.41 &0.75(Mpc)$^2\,^{(*)}$ &5 & 100 & {\sl HST} & ~\citen{Se96}\\
          &      &0.4 &1.46$\pm$0.82 & 60 &{\sl HST} & ~\citen{Sm97}\\
Cl 1358+62& 0.33 & 1&4.4  & 90$\pm$13 &{\sl HST} & ~\citen{H98}\\
Cl 1455+22& 0.259 &0.45 & 3.6 &540 & WHT$^{{\rm \ iv})}$  & ~\citen{Sm95}\\
Cl 1601+43& 0.54 &0.4 &1.54$\pm$1.32 & 95 &{\sl HST}       & ~\citen{Sm97}\\
MS 1054-03& 0.83 &1 &28$\pm$6 & 395$\pm$85 &UH$^{{\rm \  v})}$& ~\citen{LK96}\\
MS 1137+66& 0.783 &1 &4.9$\pm$1.6 & 135$\pm$50 &Keck II & ~\citen{Cl98}\\
MS 1224+20& 0.33 &0.96 &7.0& 400 &CFHT & ~\citen{F94}\\
RXJ 1347-11& 0.451 &2 &34$\pm$8 & 100$\pm$25 &CTIO & ~\citen{FT97}\\
RXJ 1716+67& 0.813 &1 &5.2$\pm$1.8 & 95$\pm$35 &Keck II & ~\citen{Cl98}\\
3C 295 & 0.46 &0.4 &4.70$\pm$0.76 & 165 &{\sl HST} & ~\citen{Sm97}\\
($=$ Cl 1409+52) & & & & & &\\
\hline
 \end{tabular}
  \end{center}
\end{table}
\noindent
{\footnotesize
{\it Notes}:\\
 a)  Redshift of the cluster.\\
 b)  Radius of the aperture 
   in units of Mpc: ($*$) area of the observed field.\\
 c)  Projected cluster mass within the aperture of radius $r$ in units of
$10^{14}M_{\odot}$.\\
 d)  Mass-to-light ratio of the cluster in
 solar units. \\
 e)  Telescopes: i) Cerro Tololo Inter-American Observatory telescope; 
 ii) Canada-France-Hawaii Telescope; iii) Hubble Space Telescope; iv)
 William Herschel Telescope; v) University of Hawaii telescope.\\
 f)  References
}
%
%
%

\clearpage

\section{Application to the cluster Abell 370}
\subsection{Background}

Abell 370 is a very rich, distant cluster of galaxies at a redshift of 0.375,
dominated by two bright cD galaxies. 
To this time, one GLA, 
several multiple images and a number of 
arclets have been observed in this cluster. The GLA was
discovered by Soucail et al.\cite{Sou87} and identified
spectroscopically as the image of a background galaxy at a redshift of
0.724 lensed by Abell 370.\cite{Sou88} \
Since the discovery of
the GLA, the cluster has been the subject of extensive
lensing studies  and
observations.\cite{Kneib93,Mel88,Sm96c,Abd98,Ota98,Bez98}  \
%
Kneib et al.\cite{Kneib93} investigated the mass distribution in Abell 370 on
the basis of their excellent ground-based CCD image.
Assuming that the GLA consists of three
merging images, they constructed a detailed mass model
based on two elliptical 
components associated with the two bright cD galaxies
by fitting the model to the observed GLA and multiple images.
Smail et al.\cite{Sm96c} 
discovered a radial-arc candidate in the {\sl HST}/WFC1 image.
Using the Kneib et al. model, they  predicted its redshift to be
$1.3\pm 0.2$.
Ota et al.\cite{Ota98} modeled the projected mass distribution in Abell 370 
on the basis of the {\sl ASCA} observation,
assuming spherical mass profiles.
They compared the radial profile of the projected mass inferred
from X-ray analysis
with that inferred from the Kneib et al. model based on strong lensing. 
The lensing mass interior to the GLA radius of $160h_{50}^{-1}$ kpc 
is about three times larger than the X-ray mass. They attributed the mass
discrepancy for this cluster to the projection effect of substructures.
Recently, AbdelSalam et al.\cite{Abd98} developed a non-parametric method to
reconstruct the  cluster-mass distribution based on the
observational constraints by strong lensing. 
They applied their non-parametric 
 method to Abell 370; they divided the projected
cluster mass into squire pixels and searched for one of the solution for
the mass distribution that follows galaxy light as closely as possible.
From a visual inspection of the {\sl HST}/WFC1 image,\cite{Sm96c} \ 
AbdelSalam et al. interpreted the GLA as 
a five-image system in contrast to the Kneib et
al. model. 
Using the GLA,
the multiple images and the radial arc,
they obtained mass maps for Abell 370, which reveal an obvious bimodal
feature in the mass distribution with the two mass peaks close to the two cD 
galaxies. 
However, the resulting two mass maxima were slightly closer
to each other than the two cD galaxies. 
In addition, the southern mass
clump was found to be more massive than the northern one, though
the northern cD galaxy is brighter than the southern one, consistent
with the Kneib et al. model. 
The interesting result obtained
there is the presence of an extra substructure 
close to the two mass peaks, which does
not follow the light. B\'{e}zecourt et al.\cite{Bez98} improved the 
Kneib et al. model using the deep WFPC2 image. 
Detailed
information concerning the GLA, multiple images 
and arclets placed strong
constraints on the mass model consisting of the cluster- and
galaxy-scale components.

In this section we present a non-parametric reconstruction of the mass
distribution for Abell 370 based on weak-lensing analysis. 
In contrast to the mass reconstruction based on strong lensing as
performed by Kneib et al., AbdelSalam et al. and
B\'{e}zecourt et al., 
we do not use the information regarding the redshifts of the arcs: 
We make use of the information on the image shapes alone.
Throughout this section, we assume a matter-dominated 
 Einstein-de Sitter universe with $\Omega_0=1$.


\subsection{Observation and data reduction}

\begin{wrapfigure}{r}{6.6cm}
\epsfxsize=8 cm
\centerline{\epsfbox{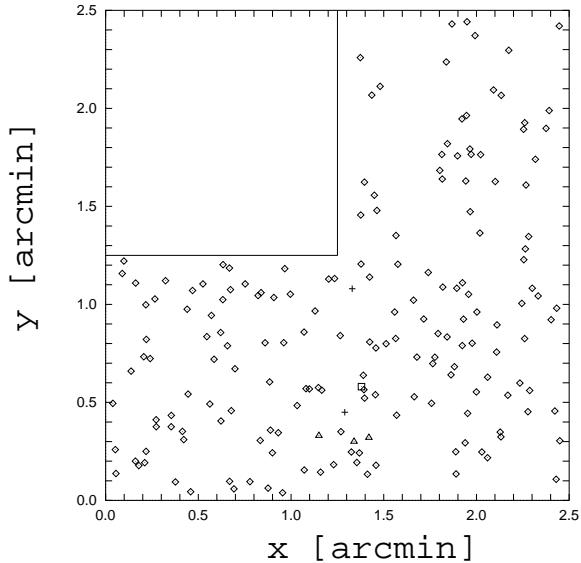}}
\caption{
Relative locations of the two cD galaxies and the 
image systems contained in our catalog of 180 galaxy images.
North is to the top, and East is to the left.
The positions of the two cD galaxies are marked with $+$. 
The positions of the giant luminous arc  
and the radial arc are marked with $\triangle$ and $\Box$,
 respectively. The positions of the other arclets 
are marked with $\Diamond$.
}
\label{location}
\end{wrapfigure}

The data for Abell 370 have been retrieved from the {\sl HST} archive.
The cluster Abell 370 was observed in December 1995
using the WFPC2 camera with the F675W filter ($R_{675W}$) on the {\sl HST} 
(ID: 6003, P.I.: R. P. Saglia).
The total exposure time is $5600$ seconds. 
After STScI pipeline processing, the data were shifted and combined into 
the final frame to remove cosmic rays and hot pixels 
using the IRAF/STSDAS task CRREJ. 
We discarded the PC chip 
from our analysis because of its brighter isophotal limit, and
thus the final frame consists of three WFPC chips.
The side length of the data field is about $2.\am 5$, corresponding to
$0.93 h_{50}^{-1}$ Mpc at redshift $z_{\rm d}=0.375$. 
($1''$ on the sky represents $6.2 h_{50}^{-1}$ kpc.)

To construct a catalog of faint objects in this frame and measure the
shape parameters (i.e., the center, the size and the complex
ellipticity $\epsilon$ defined by Eq. (\ref{Qij})) for each object, 
we used the SExtractor package.\cite{BA96} \ 
We selected all objects with isophotal areas
larger than 12 pixels (0.12 arcsec$^2$) above a detection threshold of 
$2\sigma$/pixel,
corresponding to $\mu_{\rm 675W}=24.5$ mag arcsec$^{-2}$.
The 
first and second brightness moments were computed for
each object to determine the shape parameters. 
The faint and the bright magnitude limits were chosen
so as to give reliable shape parameters. 
A catalog for arclet candidates 
was constructed with a total of 177 galaxies in a magnitude range
$R_{\rm 675W}\in 
 (23,25)$.

As mentioned above, one GLA ($z=0.724$) and one radial-arc candidate 
have been observed in the 
field of Abell 370. 
Through the detection process described above,
the radial arc was
identified as an object with $R_{\rm 675W}=24.0$, 
and thus it was included in
our catalog of faint galaxies with $R_{\rm 675W}\in  (23,25)$.
On the other hand, the GLA was identified as 
a three-component image with $R_{\rm 675W}=20.7$, $20.6$ and $20.5$,
so that this image system was excluded from our catalog of faint
galaxies with 
$R_{\rm 675W}\in  (23,25)$.  However, the GLA provides us with invaluable
information on the mass distribution
especially in the central region,
and thus it will strongly constrain the cluster profile. 
Hence, we include this image system in our galaxy catalog.
%
%
 Thus the total number of the arclets contained in our galaxy catalog
 is $N_{\rm gal}=180$, corresponding to  38  galaxies arcmin$^{-2}$. 
In Fig. \ref{location} we show the relative locations of the two cD
galaxies and the image systems contained in our catalog of 180 galaxy
images.

 
\subsection{Shear field}

\begin{wrapfigure}{r}{6.6cm}
\epsfxsize=8 cm
\centerline{\epsfbox{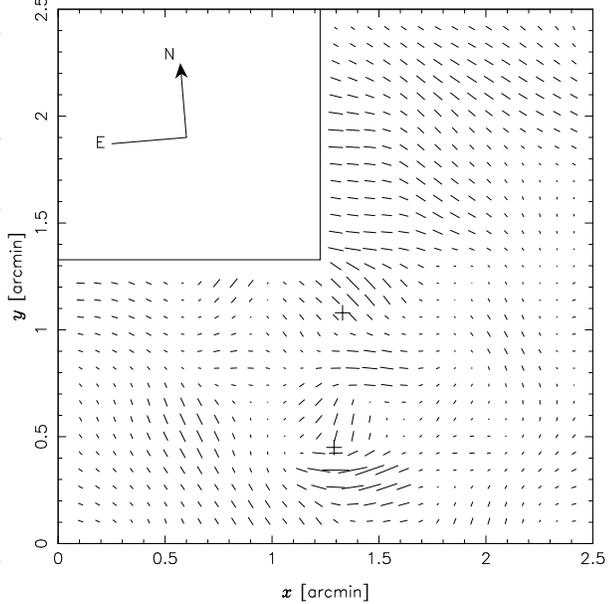}}
\caption
{
The shear field obtained from 180 galaxy images taken with
 the  {\sl HST}/WFPC2.  
The orientation and the length of a vector indicate the orientation $\phi$
 and the absolute value $|\bar{\epsilon}|$ of the locally-averaged image 
 ellipticity $\bar{\epsilon}=|\bar{\epsilon}|\,{\rm e}^{2i\phi}$,
 respectively. 
The case $|\bar{\epsilon}|=1$ (i.e., vanishing axis ratio)
corresponds to a vector of length $0.\am 3$.
We adopt an optimal smoothing length $\Delta\theta$ depending on the 
distance from the
positions 
$\vec{\theta}_{\rm N}$ and $\vec{\theta}_{\rm S}$ 
of the northern and the southern cD galaxies, respectively; 
$\vec{\theta}_{\rm N}=(1.\am 33,
1.\am 08)$ and  $\vec{\theta}_{\rm S}=(1.\am 29, 0.\am 45)$. 
The smoothing length $\Delta\theta$ ranges from 
 $\sim 0.\am 1$  ($37$ kpc) to $\sim 
0.\am 3$  ($112$ kpc), depending on the grid position;
$1'$ corresponds to $0.37$ Mpc at the cluster redshift $z_{\rm d}=0.375$
for $H_0=50$ km s$^{-1}$ Mpc$^{-1}$ and an
 Einstein-de Sitter universe.
The positions of the two cD galaxies are marked with $+$.
}
\label{shear}
\end{wrapfigure}

In this subsection we describe the shear analysis of Abell 370.  
First, we divide the CCD-data field consisting of three
WFC chips into two
rectangular fields A and B; field A with a side length of about
$2.\am 5\times 1.\am 25$  consisting of  the lower-left and the lower-right 
WFC chips, and field B with a side length of about $1.\am 25\times
2.\am 5$ consisting of the lower-right and upper-right WFC chips. 
Both the  cD 
galaxies are located at the left edge of the lower-right WFC chip (see
Fig. \ref{location}),
and thus both the fields A and B contain them. 
For field A, the number of the galaxies contained in our catalog is
142, corresponding to a galaxy-number density of 46
arcmin$^{-2}$. 
On the other hand, for field B, 
the number of galaxies contained in our catalog 
is 107, and the corresponding galaxy-number density 
is 35 arcmin$^{-2}$, 
 which is much smaller than 
%
%
a typical galaxy-number density of $\sim 50$ galaxies arcmin$^{-2}$ 
in a weak-lensing  analysis. 
This small number density is ascribed to the small number of faint
galaxies detected in the upper-right WFC frame from our analysis 
(38 galaxies in this frame; see Fig. \ref{location}).   


Next, we introduce a rectangular grid with a constant
grid separation of $0.\am 083$ ($31h_{50}^{-1}$ kpc) for each field: 
field A with $30\times 15$ grid points, and 
field B with $15\times 30$ grid points.
In order to reduce the noise due to the intrinsic ellipticities of the
background galaxies,  
 we calculate the local mean image ellipticity
 $\bar{\epsilon}(\vec{\theta}_{ij})$ at each  grid point
 $\vec{\theta}_{ij}$ 
from the image ellipticity $\epsilon(\vec{\theta}_k)$ of the $k$-th galaxy 
 at angular position $\vec{\theta}_k$ ($k=1,2,\cdots, N_{\rm gal}$;
 $N_{\rm gal}=180$) 
(see Eqs. (\ref{epsbar}) and (\ref{weightfunc})). 
%
%
 Here we employ an optimal smoothing procedure; 
we adopt an optimal
 smoothing length $\Delta\theta(\vec{\theta}_{ij})$ in Eq. (\ref{weightfunc})
depending on the grid position $\vec{\theta}_{ij}$ such that
\begin{equation}
 \Delta\theta(\vec{\theta}_{ij})=
\min(0.12\,|\vec{\theta}_{ij}-\vec{\theta}_{\rm N}|+\Delta\theta_0,\; 
     0.12\,|\vec{\theta}_{ij}-\vec{\theta}_{\rm S}|+\Delta\theta_0 )
\;\;\mbox{arcmin},
\end{equation}
with constant smoothing length 
$\Delta\theta_0=0.\am 1$,
where $\vec{\theta}_{\rm N}$ 
and $\vec{\theta}_{\rm S}$ are the angular positions of the northern and 
the southern cD galaxies, respectively; $\vec{\theta}_{\rm N}=(1.\am 33,
1.\am 08)$ and  $\vec{\theta}_{\rm S}=(1.\am 29, 0.\am 45)$ 
in our coordinate system. 
 Thus the smoothing length
$\Delta\theta(\vec{\theta}_{ij})$ at grid position $\vec{\theta}_{ij}$ 
ranges from $\sim 0.\am 1$  ($37 h_{50}^{-1}$ kpc) to $\sim 
0.\am 3$  ($112 h_{50}^{-1}$ kpc), 
depending on the distance from the positions $\vec{\theta}_{\rm N}$ and
$\vec{\theta}_{\rm S}$ of the two cD galaxies.

Figure \ref{shear} displays the resulting map of the locally-averaged image 
ellipticities 
$\bar{\epsilon}=|\bar{\epsilon}|\,{\rm
e}^{2i\phi}$ obtained using 180 arclet candidates. The orientation of a
vector indicates the direction $\phi$ of the shear, and the length of a
vector is proportional to the strength of the shear.
A coherent shear pattern induced by the  
gravitational field of the cluster
is clearly visible.


\subsection{Mass reconstruction}

Using the smoothed image-ellipticity field
$\bar{\epsilon}(\vec{\theta}_{ij})$ obtained from our galaxy catalog, 
we reconstruct the mass
distribution of the cluster Abell 370. 
To take into account (1) the strong-lensing features in the cluster field, 
(2) the fairly high redshift
of the cluster ($z_{\rm d}=0.375$) and (3) the small field-of-view of the 
{\sl HST}/WFPC2 field, we follow
%
the generalized
mass-reconstruction scheme\cite{SeSc97} 
based on the finite-field inversion derived by Seitz and
Schneider\cite{SeSc96} (see {\S} 4.3 and 4.4). 

\begin{wrapfigure}{r}{6.6cm}
\epsfxsize=5 cm
\centerline{\epsfbox{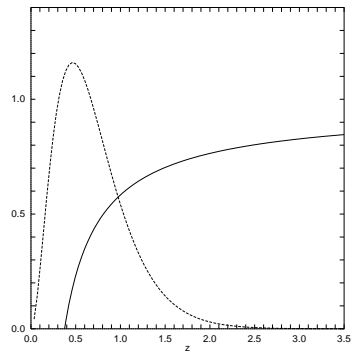}}
\caption{
The solid line indicates the relative lensing strength
 $w(z)$ defined by Eq. (\protect\ref{wzEde}) 
 for the cluster redshift $z_{\rm d}=0.375$.
The dashed line indicates
the assumed redshift distribution $p_z(z)$ of background
 galaxies defined by Eq. (\protect \ref{pz}) for $\beta=1$ and
 $\bk{z}=3z_0=0.7$. 
}
\label{red}
\end{wrapfigure}

Because of the irregular shape of the WFC field, we 
separately reconstruct  the
two-dimensional convergence fields $\kappa_{\infty,{\rm
A}}(\vec{\theta}_{ij})$ and $\kappa_{\infty,{\rm B}}(\vec{\theta}_{ij})$
on two rectangular fields A and B, respectively, as done in
Ref.~\citen{Se96}.  
Since a mass reconstruction with the finite-field formula
(\ref{KernelH2}) yields a constant $\bar{\kappa}_{\infty}$, 
which represents the
average of $\kappa_{\infty}$ over the data field, 
we have two free
constants in our reconstruction, 
$\bar{\kappa}_{\infty,{\rm A}}$ for field A and 
$\bar{\kappa}_{\infty,{\rm B}}$ for field B. 
One of the two constants can be used to join the two
reconstructions, and 
the residual constant can be used 
to normalize the cluster mass.
The procedure to determine the two constants and to join the two independent 
reconstructions is described in {\S} 6.5.

Moreover, we must know the 
redshift distribution of the
galaxy population in the mass reconstruction (see {\S} 4.4). 
However, the redshift distribution of faint galaxies with
$R_{\rm 675W}\in(23,25)$  is uncertain, so that
%
we adopt a parameterized redshift distribution $p_z(z)$ 
of field galaxies of the form\cite{Se96,Br96}  
\begin{equation}
p_z(z)=\frac{\beta z^2}{\Gamma(3/\beta)z_0^3}\exp
\left(
-(z/z_0)^{\beta}
\right),\label{pz}
\end{equation}
with mean redshift
$\bk{z}=z_0\Gamma(4/\beta)/\Gamma(3/\beta)$.
We use the redshift distribution (\ref{pz})
with $\beta=1$ and $\bk{z}=3z_0=0.7$ in our analysis.
In Fig. \ref{red} we show this redshift distribution $p_z(z)$ for 
$\beta=1$ and $\bk{z}=3z_0=0.7$ and the relative lensing strength
$w(z)$ defined by Eq. (\ref{wzEde}) 
for the cluster redshift $z_{\rm d}=0.375$.


\subsection{Results}

\begin{figure}[htb]
\parbox{\halftext}{
\epsfxsize=6.8 cm
\centerline{\epsfbox{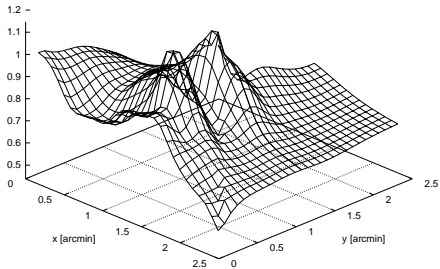}}
\caption{The $\kappa_{\infty}$-distribution for the cluster Abell
 370 ($z_{\rm d}=0.375$) reconstructed from the shear field shown in
 Fig. \ref{shear}.  
 The side length is about $2.\am 5$
 corresponding to 0.93 Mpc at the cluster redshift $z_{\rm d}=0.375$ 
for $H_0=50$ km s$^{-1}$ Mpc$^{-1}$ and an
 Einstein-de Sitter universe.
In the reconstruction we used the redshift
 distribution $p_z(z)$ defined by Eq. (\protect\ref{pz}) with $\beta=1$ and
 $\bk{z}=0.7$.   
Our reconstruction exhibits a clear bimodal feature in the central region
 of the cluster. A likely substructure is visible near to the two cD galaxies.
}
\label{k3d}}
\hspace{8mm}
\parbox{\halftext}{
\epsfxsize=6.5 cm
\centerline{\epsfbox{fig6.eps}}
\caption
{
Contour plot of the resulting $\kappa_{\infty}$-distribution shown in
 Fig. \ref{k3d}.
The positions of the two cD galaxies are marked with $+$.
The position of the northern mass clump agrees well with that of the
 northern cD galaxy. The southern mass clump is located slightly offset
 from the southern cD galaxy. The position of the extra substructure is
 marked with $\star$.  The contours are stepped in units of 0.05.
}
\label{k2d}}
\end{figure}

Figure \ref{k3d} displays the two-dimensional 
$\kappa_{\infty}$-field for Abell 370 obtained 
using our galaxy sample and
 the redshift distribution (\ref{pz}) with $\beta=1$ and $\bk{z}=0.7$.
In Fig. \ref{k2d} we show the corresponding contour map of the
$\kappa_{\infty}$-reconstruction.

In the reconstruction, 
 we have determined the two constants $\bar{\kappa}_{\infty,{\rm
A}}$ and $\bar{\kappa}_{\infty,{\rm B}}$ in the following way:
(i) We employ the well-constrained
mass model constructed by
Kneib et al.\cite{Kneib93} using the GLA and multiple images. 
The mean convergence within field A is calculated to be 0.832 using this 
model. 
(ii) Adopting this value for $\bar{\kappa}_{\infty,{\rm A}}$,
we determine $\bar{\kappa}_{\infty,{\rm B}}$ such that the averages of
$\kappa_{\infty,{\rm A}}(\vec{\theta}_{ij})$ and 
$\kappa_{\infty,{\rm B}}(\vec{\theta}_{ij})$  
within the overlapping region of
the fields A and B are the same. The mean convergence in this
overlapping region is 0.786, and the mean convergence in field B is 0.764. 
%
(iii) Joining together the two independent $\kappa_{\infty}$-reconstructions
along the diagonal of the overlapping quadrate, we obtain the resulting
convergence map $\kappa_{\infty}(\vec{\theta}_{ij})$,
as was done in Ref~\citen{Se96}. 
It can be seen in Figs. \ref{k3d} and \ref{k2d} 
that the discontinuity across this
diagonal in our reconstruction is remarkably small.
From the resulting $\bar{\kappa}_{\infty}$-map, 
the mean convergence $\bar{\kappa}_{\infty}$ inside the observed field 
of $0.65 h_{50}^{-2}$ Mpc$^2$ is calculated to be 0.811, yielding a
total mass in this field of $M\simeq 8.0\times 10^{14} h_{50}^{-1}
M_{\odot}$. 

The main features in the resulting mass map are summarized as follows:\\
(1) Our mass reconstruction exhibits a clear bimodal feature in the
central region where the two cD galaxies are located. The locations of
the two mass
maxima ($\kappa_{\infty}=1.14$)
in the mass map coincide well with those of 
the two cD galaxies: 
The location of the northern mass clump is in good agreement 
with that of the northern cD galaxy, while
the southern mass clump is located slightly offset from the southern cD
galaxy, but is consistent with the location of the
southern cD galaxy within a smoothing scale and a grid separation of $\sim
0.\am 1$ ($37h_{50}^{-1}$ kpc). \\
(2) The reconstructed mass map reveals a likely substructure (marked
with $\star$ in Fig. \ref{k2d})
located about $0.\am 3$ ($110$ kpc) to the left relative to the center
of the two cD galaxies.\\
%
%
(3) An increase of $\kappa_{\infty}$ toward the lower-left corner in
the lower-left quadrate is visible.\\ 
%
%
(4) The cluster-mass distribution is super-critical especially 
in the innermost region where the two cD galaxies and hence 
the two mass clumps lie. To be more 
precise, this distribution is
super-critical for sources at an infinite or a sufficiently high 
redshift. However, the reconstructed mass distribution is sub-critical
for the GLA at a redshift of 0.724; $w(z)=0.45$ for $z=0.724$ 
(see Eq. (\ref{wzEde})).

\vspace{2mm}
\noindent
Finally, it should be mentioned that these main features are insensitive to
the assumed redshift distribution $p_z(z)$ of field galaxies and the
smoothing procedure which is needed in calculating the smoothed image
ellipticity $\bar{\epsilon}$.

%
%
%


\subsection{Discussion}

Focusing on 
the morphology of the cluster, 
we have investigated the mass 
distribution of 
Abell 370 on the basis of weak lensing.
We reconstructed the two-dimensional  mass distribution of 
Abell 370 from the ellipticities of the faint galaxy images  
obtained using the {\sl HST}/WFPC2 data. 
In the reconstruction, we used a total of 180 galaxy images consisting
of 176 arclets, a radial arc and a GLA as a three-image system.
Taking account of the small field-of-view of the data field, the
strong-lensing 
features in the cluster and the fairly high redshift ($z_{\rm d}=0.375$) 
of the cluster, we applied 
the generalized mass-reconstruction scheme described in Section 4.4. 
From the shear field of galaxy images, 
the mass distribution of the cluster was determined up to a 
one-parameter family of the global transformations
which leaves the observed image distortions
invariant.
We employed the well-constrained mass model by Kneib et
al.\cite{Kneib93} based 
on strong lensing in order to infer the mass in the central region,
from which we broke the mass degeneracy.
Assuming a mean redshift of
$\bk{z}=0.7$ for the field galaxies with $R_{\rm 
675W}\in (23,25)$, 
we have estimated  the cluster mass within our data field of $0.65
h_{50}^{-2}$ Mpc$^2$ to be $\sim 8\times 10^{14} h_{50}^{-1} M_{\odot}$.

The resulting mass map  also provides valuable information
regarding the distribution of dark matter.
Our direct mass reconstruction exhibits 
a clear bimodal feature associated with the two cD galaxies. Furthermore,
our mass reconstruction reveals some other features: An extra
substructure in the vicinity of  
the two cD galaxies and a mass condensation toward the 
lower-left boundary are visible.
However, since the accuracy of mass reconstructions tends to be 
worse near the
boundary of the data field, the mass increase toward the boundary
corner might be due to the
systematic 
boundary effect and/or the fact that fewer galaxy images are used at the
boundary corner in averaging image ellipticities. 
In addition, since the galaxy-number density  in the upper-right WFC
is quite small (see Fig. \ref{location}), the reconstructed  mass
distribution  may be different from the actual one in that region.  
The resulting mass distribution of the cluster is found to be
super-critical for
sources at a sufficiently-high redshift but sub-critical for the GLA at
a redshift of 0.724. This result can be attributed to the smoothing
procedure which is needed to reduce the noise due to the intrinsic
ellipticities of background galaxies:
Although the global mass distribution of a cluster
can be obtained by locally averaging the
galaxy-image ellipticities,
such averaging will smooth
out the galaxy-scale structure in the cluster.
As a result, the reconstructed mass
distribution will tend to be flatter than the original one. 
Hence, strong-lensing effects such as GLAs and multiple images
should be combined with weak-lensing analysis in order
to reproduce the mass
distribution in the innermost region of the cluster.
Alternatively, the galaxy-galaxy lensing analysis can probe the mass
distributions in clusters on galaxy scales.\cite{NK97,Nat98,GS98,GS99} 

Finally, we should comment on our treatment of arcs: a GLA and
a radial arc. 
In general, strong-lensing features such as GLAs and radial arcs
strongly constrain the mass distribution of the deflector,
especially in its central region where they are located. 
In fact, the reconstruction without taking into account the arcs 
exhibits a bimodal feature, but the density of the southern peak
turns out to be lower than that of the northern peak.
Thus the comparison between the resulting mass maps 
with and without the arcs reveals
a remarkable influence of the arcs
on the reconstructed mass distribution.


\section{Summary}

Weakly- and coherently-distorted  images of background galaxies 
induced by intervening 
galaxy clusters can be used to determine the mass distributions in
the clusters on scales $\sim$ 1 Mpc. 
Kaiser and Squires have derived an exact expression for the
projected-mass distribution of the lensing cluster in terms of the
gravitational shear which can be measured from the observed galaxy images.
The original Kaiser and Squires technique was then generalized to include 
the strong-lensing regime and to optimize real observations.

To this time, 
these direct mass-reconstruction methods based on shear
analysis have been applied to a number of clusters.
Now, the weak-distortion fields of galaxy images
have become one of the most 
promising tools 
to probe the mass distribution of clusters.
Moreover, the weak-shear field also provides a clue to the distant
population of faint background galaxies:
The weak-lensing analysis for a high-redshift cluster 
combined with a strongly-constrained
mass model can  place strong constraints on the redshift
distribution of a distant population 
since the lensing
strength depends quite strongly on the source redshift through the
distance ratio $D_{\rm ds}/D_{\rm s}$ (or $w(z)$) for
a high-redshift cluster.
The detection of significant lensing signals for some high-redshift
clusters ($z_{\rm d}\sim 0.8$) indicates that a substantial part of the
faint galaxy population  must lie at sufficiently high redshifts
($z>1$). 

In this paper we have applied a direct mass-reconstruction method to the
cluster Abell 370 at a redshift of $z_{\rm d}=0.375$ observed with the
{\sl HST}/WFPC2.
Despite the small field-of-view and the irregular shape of the data
field, 
our direct mass reconstruction of Abell 370 demonstrates the feasibility
of weak-lensing analysis based on the shear field. 
However, there remains an uncertainty concerning the mass normalization of 
the cluster. Such an uncertainty 
occurs if 
the mass
distribution of the cluster extends beyond the observed data field, in
which case we cannot determine the cluster-mass distribution directly
from the observed image distortions. 
We expect that such a difficulty will be overcome by 8-10 m class
telescopes with a wide field-of-view, such as the
8.3 m Subaru telescope with Suprime-Cam.

%
%
%
%
%
%
%
\section*{Acknowledgements}
We are very grateful to P. Schneider for informing us of his
unpublished works and for his valuable suggestions.
We also wish to thank M. Hattori for  fruitful discussions concerning 
previous works on the cluster Abell 370.
Part of this work is based on an observation with the NASA/ESA {\it Hubble}
{\it Space} {\it Telescope}, obtained from the data archive at the Space 
Telescope Science Institute (STScI). STScI is operated by the
Association of Universities for Research in Astronomy, Inc. under the
NASA contract NAS 5-26555.


\end{document}